\documentclass[aip,reprint]{revtex4-2}

\usepackage{graphicx}
\usepackage{bm}
\usepackage{dcolumn}
\usepackage{physics} 
\usepackage[hidelinks]{hyperref}
\usepackage{cleveref}
\usepackage{bbm}
\usepackage{color}

\usepackage{amssymb}

\newcommand\exval[2][]{#1\left\langle {#2} #1\right\rangle}

\begin{document}


\title{Expansion of time-convolutionless non-Markovian quantum master equations:  
A case study using the Fano-Anderson model}



\author{Tim Alh\"auser}
\affiliation{Institute of Physics, University of Freiburg, Hermann-Herder-Strasse 3, D-79104 Freiburg, Germany}

\author{Heinz-Peter Breuer}
\email{breuer@physik.uni-freiburg.de}
\affiliation{Institute of Physics, University of Freiburg, Hermann-Herder-Strasse 3, D-79104 Freiburg, Germany}
\affiliation{EUCOR Centre for Quantum Science and Quantum Computing, University of Freiburg, Hermann-Herder-Str. 3, D-79104 Freiburg, Germany}

\date{\today}


\begin{abstract} \label{sec: Abstract}
	
We explore the performance of the time-convolutionless (TCL) projection operator technique using the Fano-Anderson 
model as a test case. Comparing the exact TCL master equation with an expansion in powers of the strength of the 
system-environment coupling, we analyze the transient dynamics as well as the steady-state behavior. 
For a Lorentzian spectral density we demonstrate that the dimensionless expansion parameter corresponds to the ratio of 
the environmental correlation time to the relaxation time of the system, and we derive the convergence radius for the TCL 
expansion, which is seen to depend on the ratio of detuning and width of the spectral density. We further study the 
quantum non-Markovianity of the model based on the evolution of the Bures distance between quantum states and how it is
represented by the second and fourth order of the expansion. Our results highlight both the strengths and the limitations of 
the TCL formalism in capturing key features of open quantum systems and, in particular, the challenges of 
accurately describing strongly coupled systems and non-Markovian dynamics.
	
\end{abstract}


\maketitle

\section{Introduction}

The study of open quantum systems \cite{BreuerBook} lies at the heart of modern quantum science, with applications 
ranging from quantum information processing to quantum transport and quantum thermodynamics 
\cite{ReviewCleri,Kosloff, Esposito, Stasberg, Binder2018,Gemmer2004,Donarini2024}. 
Whenever a system interacts with its 
environment, its dynamics deviates from unitary evolution, giving rise to decoherence and dissipation. A central challenge 
is thus to describe these non-unitary processes in a mathematically and physically consistent and efficient way.

Over the years, a variety of master equation approaches have been developed. Among them, time-local master equations 
obtained from the time-convolutionless (TCL) projection operator technique \cite{TCL_Shibata,TCL_Shibata2} have been 
shown to be especially useful, as they provide a systematic perturbative expansion in the system–environment interaction 
while retaining a time-local structure. As exact expressions are typically difficult to obtain, one often utilizes
a perturbation expansion in the strength of the system-environment interaction which corresponds to an expansion
in terms of ordered cumulants.\cite{vanKampen1974a,vanKampen1974b,Kampen,BreuerBook} Recently, we have
developed a recursive perturbation scheme for the determination of the TCL master equation.\cite{Colla2025b,Colla2025c}

However, the validity of expansions depends critically on the strength of the system-environment coupling, and what 
qualifies as “strong” coupling is often not uniquely defined. In this context the study of analytically solvable
system-environment models is of great interest as they allow to analyze the performance of the TCL expansion
by comparison with the exact master equation of the exact solution. Such an analysis has been carried out for
various microscopic models, e.g. for a spin-star model, \cite{Breuer2004} for the Caldeira-Leggett model \cite{Breuer2001} 
and for the model for a qubit in a cavity.\cite{Lidar} Recently a detailed study of the performance of the expansion
for a non-integrable case, the paradigmatic spin-boson model, has been carried out.\cite{Kumar2026}
Here, we investigate this issue by using the Fano-Anderson model, an analytically solvable Gaussian 
model,\cite{Fano,Anderson,Cohen_Tannoudji} which provides an example of a continuous variable system (harmonic 
oscillator) coupled to a bath with structured spectral density. As the exact master equation is known,
\cite{Tu2008,Jin2010,Lei2012,Zhang2012,AleReservoirs,Irene} 
the TCL expansion can relatively easily be obtained by an expansion of the coefficients of the master equation. As we will 
see this allows to identify the expansion parameter and to determine the radius of convergence of the expansion.

As a further important feature we study the degree of non-Markovianity (memory effects) which is
characterized by the backflow of information from the environment to the open system.\cite{Breuer2009,Breuer2016a}
We will employ the Bures distance between quantum states \cite{Hayashi} to quantify the information backflow and
the size of memory effects\cite{Einsiedler2020,Campbell,Megier,Vasile,deVega} both for the exact and the approximate
evolution. This enables us to assess to which extend the expansion can capture non-Markovian effects of the open system 
dynamics.

The paper is organized as follows. \Cref{sec:DoOS} reviews the general framework of open quantum systems and the TCL 
master equation, and introduces the Bures distance as a measure of non-Markovianity. In Sec.~\ref{sec:Framework} we 
present the Fano-Anderson model and outline the perturbative expansion procedure. In Sec.~\ref{sec:Results}, we provide 
both analytical and numerical results, comparing the perturbatively generated dynamics with the exact solution. 
\Cref{sec:NM} investigates the emergence of non-Markovianity in the model and evaluates the capability of the perturbative 
treatment to capture information backflow effects. Finally, in Sec.~\ref{sec:Conclusion} we summarize the main findings and 
discuss their implications for strongly coupled open quantum systems.


\section{Theoretical Background} \label{sec:DoOS}

This section briefly summarizes the description of the dynamics of open quantum systems via the TCL master equation. 
We introduce the general formalism of the reduced system dynamics, the perturbative expansion of the TCL generator, 
and the quantification of non-Markovian effects using the Bures distance. 

\subsection{Open Quantum Systems and TCL Master Equations} \label{sec:OS}

In the description of open quantum systems one typically considers a system of interest interacting with an environment. 
The dynamics of the total closed system is unitary and governed by the von~Neumann equation for the density matrix
$\rho$,\cite{BreuerBook}
\begin{align}
	\dot{\rho}(t) = -i\hbar\left[H_{SE}(t),\,\rho(t)\right], \label{eq:vonNeumann}
\end{align}
with a total Hamiltonian composed of system, environment, and interaction contributions,
\begin{align}
	H_{SE}(t) = H_S(t)\otimes\mathbbm{1}_E 
	+ \mathbbm{1}_S\otimes H_E(t) 
	+ H_I(t).
\end{align}
We adopt units where $\hbar=1$ so that all energies and frequencies are expressed in the same units.
We assume that at the initial time $t=0$ system and environment are uncorrelated,
\begin{align}
	\rho(0) = \rho_S(0)\otimes\rho_E(0) , \label{eq:factorized_init}
\end{align}
where the initial environmental state is supposed to be a thermal state
\begin{align}
	\rho_E(0) = \frac{\exp(-\beta H_E)}{Z}
\end{align}
with the inverse environmental temperature $\beta$ and the partition function~$Z = \Tr_E\{\exp(-\beta H_E)\}$, ensuring 
that the environment is initially in thermal equilibrium at temperature $T = 1/\beta$.

Since tracking all environmental degrees of freedom is generally infeasible, particularly for large or even infinite baths, 
the description is reduced to the system degrees of freedom by tracing out the environment,
\begin{align}
	\rho_S(t) = \Tr_E \left\{\rho(t)\right\}.
\end{align}
The time evolution of the reduced state can be expressed by a dynamical map $\Phi_t$ that connects the initial and 
time-evolved system states,
\begin{align}\label{eq:Rhot}
	\rho_S(t) = \Phi_t\left[\rho_S(0)\right].
\end{align}
Differentiating this relation and substituting the initial state from \eqref{eq:Rhot} yields a time-convolutionless master 
equation of the form
\begin{align} \label{TCL-master}
	\dot{\rho}_S(t) = \dot{\Phi}_t\Phi^{-1}_t\left[\rho_S(t)\right] 
	= \mathcal{L}_t\!\left[\rho_S(t)\right].
\end{align}
This is a first-order differential equation which is local in time with a generator $\mathcal{L}_t$, also known as TCL
generator, which generally depends explicitly on time $t$. The TCL formalism thus provides a compact, time-local 
description of the system dynamics without explicit reference to the past history of the system.

The TCL generator $\mathcal{L}_t$ can always be written in a form analogous to the Lindblad form,\cite{BreuerQNM,Andersson2014a}
\begin{align}
	\mathcal{L}_t\!\left[\rho_S(t)\right] 
	= -i\!\left[K_S(t), \rho_S(t)\right] 
	+ \mathcal{D}_t\!\left[\rho_S(t)\right],
\end{align}
where the first term represents coherent Hamiltonian evolution and the dissipator $\mathcal{D}_t\!\left[\rho_S(t)\right]$,
accounting for dissipation and decoherence, can be written as
\begin{equation}
	\mathcal{D}_t\!\left[\rho_S\right] = \sum_k \gamma_k(t) 
	\left[ L_k(t)\rho_SL_k^\dagger(t) - \frac{1}{2}\!\left\{L_k^\dagger(t)L_k(t), \rho_S\right\} \right]
\end{equation}
with time-dependent Lindblad operators $L_k(t)$ and rates $\gamma_k(t)$. In the non-Markovian regime, the rates may 
temporarily become negative, reflecting information backflow between system and environment.\cite{Breuer1999b,NonMark,Breuer2016a}



\subsection{Expanding the TCL Generator} \label{sec:TheroyExpansion}

The time-convolutionless projection operator technique\cite{TCL_Shibata,TCL_Shibata2} leads to an expansion of the 
generator $\mathcal{L}_t$ of the TCL master equation  \eqref{TCL-master} in powers of the strength of the 
system-environment coupling,
\begin{align}\label{eq:ExpansionGenerator}
	\mathcal{L}_t\left[\rho_S(t)\right] = \sum_{n=0}^{\infty} \mathcal{L}^{(n)}_t\left[\rho_S(t)\right],
\end{align}
where $\mathcal{L}^{(n)}_t$ denotes the contribution of $n$th order in the coupling. It is well known that this
expansion corresponds to an expansion in terms of ordered cumulants of the Liouville superoperator describing the
system-environment interaction.\cite{vanKampen1974a,vanKampen1974b,Breuer2001,BreuerBook}

However, in Sec.~\ref{TCL-Expansion-FA-Model} we will obtain this expansion by means of a simpler method. 
In fact, as we know the exact master equation of the Fano-Anderson model studied here, we can directly find the
expansion \eqref{eq:ExpansionGenerator} from a Taylor expansion of the time-dependent c-number coefficients 
appearing in the master equation [see Eq.~\eqref{eq:FAME}]. It is clear that both methods lead to one and the
same power expansion of the generator.\cite{BreuerBook}


\subsection{Quantum Non-Markovianity} \label{sec:TheoryNM}

In this paper we employ the characterization of non-Markovianity as a backflow of information from the
environment to the open system, expressing the presence of memory effects.\cite{NonMark,NonMark2} 
This backflow of information is signified by an increase in time of the distance between two quantum states $\rho_1(t)$
and $\rho_2(t)$ of the open system. 
While originally the trace distance was used as a suitable measure for the distance of quantum states, 
here we will use the Bures distance\cite{Hayashi} $D_B\left(\rho_1(t), \rho_2(t)\right)$, which is more appropriate for 
continuous variable systems and Gaussian states.\cite{Einsiedler2020,Vasile}
Thus, we define a measure $\mathcal{N}$ for non-Markovianity which is based on the temporal increase of the Bures 
distance by means of:
\begin{equation}\label{eq:NM}
	\mathcal{N} = \int_{\sigma > 0} d\tau \, \sigma(\tau), 
\end{equation}
where
\begin{equation}	
	\quad \sigma(t) = \frac{d}{dt}D_B\left(\rho_1(t),\rho_2(t)\right).
\end{equation}
The integral in Eq.~\eqref{eq:NM} quantifies the total amount of information flowing back from the environment to the 
system, thus measuring the total degree of non-Markovian effects.

Since Gaussian quantum states\cite{Serafini2017} are completely characterized by their first and second moments, the 
fidelity between two such states can be evaluated analytically:
\begin{align}\label{eq:Fidelity}
	\mathcal{F}\left(\rho_1, \rho_2\right) = \frac{2}{\sqrt{\Delta + \delta} - \sqrt{\delta}} e^{-\frac{1}{2}\left(\boldsymbol{d}_2 - 
	\boldsymbol{d}_1\right)^T \left(\sigma_1 + \sigma_2\right)^{-1} \left(\boldsymbol{d}_2 - \boldsymbol{d}_1\right) },
\end{align}
with 
\begin{align}
	\Delta &= 4 \det\left(\sigma_1 + \sigma_2\right), \\
	\delta &= 16\left(\det\left(\sigma_1\right) - \frac{1}{4}\right)\left(\det\left(\sigma_2\right) - \frac{1}{4}\right) .
\end{align}
The matrices $\sigma_{1}$ and $\sigma_{2}$ are the covariance matrices of the respective states, 
with each matrix given by
\begin{align}
	\sigma_{1/2} = \begin{pmatrix}
		\langle X^2 \rangle - \langle X \rangle^2 \quad\quad \frac{1}{2} \langle X P + P X \rangle - \langle X \rangle \langle P \rangle \\
		\frac{1}{2} \langle X P + P X \rangle - \langle X \rangle \langle P \rangle \quad\quad \langle P^2 \rangle - \langle P \rangle^2
	\end{pmatrix},
\end{align}
where the operators $X$ and $P$ are the usual dimensionless quadratures, defined as 
$X = \frac{\hat{a} + \hat{a}^\dagger}{\sqrt{2}}$, and  $P = \frac{\hat{a} - \hat{a}^\dagger}{\sqrt{2}i}$.
The displacement vectors $\boldsymbol{d}_{1}$ and $\boldsymbol{d}_{2}$ contain the first moments of the quadratures 
and are given by
\begin{align}
	\boldsymbol{d}_{1/2} =
	\begin{pmatrix}
		\langle X \rangle \\
		\langle P \rangle
	\end{pmatrix}_{1/2}.
\end{align}
These vectors describe the mean position of each Gaussian state in phase space, while the covariance matrices 
$\sigma_{1/2}$ characterize their shape and orientation. The covariance matrices together with the displacement 
vectors fully determine the Gaussian states $\rho_1$ and $\rho_2$. Using these quantities, the fidelity \eqref{eq:Fidelity} 
and hence the Bures distance between two states, which is given by
\begin{align} \label{Bures-dist}
	D_B\left(\rho_1, \rho_2\right) = \sqrt{2-2\sqrt{\mathcal{F}\left(\rho_1, \rho_2\right)}},
\end{align}
can easily be evaluated analytically.


\section{Application to the Fano-Anderson model} \label{sec:Framework}

Now we apply the general formalism to the analytically solvable Fano-Anderson model, which serves as a benchmark for 
testing the TCL expansion. This section introduces the model, the exact master equation, the spectral density, and 
develops the TCL expansion of the master equation.

\subsection{Fano-Anderson Model} \label{sec:FA}

The Fano-Anderson model\cite{Fano, Anderson} is particularly well suited for applications to quantum transport scenarios 
and impurities in solids. It can be viewed as the Caldeira-Leggett model~\cite{Caldeira} under the application of 
the rotating wave approximation, neglecting the fast oscillating terms in the Hamiltonian. The Hamiltonian of the 
Fano-Anderson model is given by
\begin{equation}\label{eq:H_SE}
	H_{SE} = \omega_0 a^\dagger a + \sum_k \omega_k b_k^\dagger b_k 
	+ \sum_k \left( \alpha_k a^\dagger b_k + \alpha_k^\ast a b_k^\dagger \right),
\end{equation}
describing a system oscillator with bare frequency $\omega_0$ coupled to a bath of environmental oscillators with 
frequencies $\omega_k$. These bosonic modes of system and environment are coupled by the exchange of excitations 
with coupling strengths given by $\alpha_k$. If the system loses an excitation it is absorbed by an environmental mode or 
vice versa. By this structure it is ensured that the total excitation number of the closed system is conserved. 

The Heisenberg equations of motions for this model lead to an integro-differential equation for the system mode $a$:
\begin{align}\label{eq:DEQSysMode}
	\frac{d}{dt}a(t)+i\omega_0a(t)+\int_0^td\tau\mathcal{K}(t-\tau)a(\tau) = b(t), 
\end{align}
where $b(t)$ is given by
\begin{align}
	b(t) = -i \sum_k \alpha_k \exp(-i\omega_kt)b_k(0) ,
\end{align}
and we have defined the memory kernel
\begin{align}
	\mathcal{K}(t)=\sum_k\abs{\alpha_k}^2\exp(-i\omega_kt).
\end{align}
One can rewrite the memory kernel as $\mathcal{K}(t)=\int_0^{\infty} d\omega J(\omega)\exp(-i\omega t)$ by introducing 
the spectral density
\begin{align}
	J(\omega)=\sum_k \abs{\alpha_k}^2\delta(\omega-\omega_k).
\end{align}
To represent the exact master equation it is convenient to introduce the Green function $G(t)$
which is determined by the homogeneous part of the integro-differential equation \eqref{eq:DEQSysMode},
\begin{align}\label{eq:IDE}
	\dot{G}(t) + i\omega_0 G(t) + \int_0^t d\tau \, \mathcal{K}(t-\tau)\, G(\tau) = 0 ,
\end{align}
together with the initial condition $G(0)=1$. 
We note that the function defined by $\hat{G}(t)=\theta(t)G(t)$ 
satisfies equation \eqref{eq:IDE} with a delta function $\delta(t)$ replacing the zero on the right-hand side.


\subsection{Exact Master Equation}

The exact TCL master equation for the reduced density matrix describing the state of mode $a$ is given 
by\cite{AleOG,Irene,Zhang}
\begin{align}\label{eq:FAME}
	\frac{d}{dt}\rho_S(t) &= -i\left[\omega_r(t)a^\dagger a, \rho_S(t)\right] \nonumber \\
	&+ \underbrace{\gamma(t)\left(N(t)+1\right)}_{=\gamma_-(t)} \left[a\rho_S(t)a^\dagger-\frac{1}{2}\left\{ a^\dagger a, 
	\rho_S(t) \right\}\right] \nonumber \\
	&+ \underbrace{\gamma(t)N(t)}_{=\gamma_{+}(t)} \left[a^\dagger\rho_S(t)a-\frac{1}{2}\left\{ aa^\dagger, \rho_S(t) \right\}\right] ,
\end{align}
with a time-dependent renormalized frequency $\omega_r(t)$ of the system oscillator, an emission/loss rate 
$\gamma_-(t)$, and absorption/gain contribution with rate $\gamma_+(t)$, while $N(t)$ represents an 
effective thermal occupation number.
These time dependent coefficients of the master equation can be expressed explicitly as  
\begin{align}
	\omega_r(t) &= -\,\mathrm{Im}\!\left[\frac{\dot{G}(t)}{G(t)}\right], \label{eq:CoeffExactwr} \\
	\gamma(t)   &= -\,2\,\mathrm{Re}\!\left[\frac{\dot{G}(t)}{G(t)}\right], \label{eq:CoeffExactGam} \\
	\gamma_{+}(t) &= \gamma(t)\,\mathcal{I}(t) + \dot{\mathcal{I}}(t) \label{eq:CoeffExactGamP} ,
\end{align}
where the noise integral is defined by  
\begin{align}\label{eq:NoiseInt}
	\mathcal{I}(t) = \int_0^\infty \! d\omega \, J(\omega)\, n(\omega)\,
	\left|\int_0^t \! d\tau \, G(\tau) e^{i\omega\tau}\right|^2 .
\end{align}
Here, $n(\omega)=\big(\mathrm{e}^{\beta\omega}-1\big)^{-1}$ denotes the Bose-Einstein distribution 
at inverse temperature $\beta$.


\subsection{Lorentzian Spectral Density} \label{sec:LSD}

We assume that the environmental frequency distribution is described by a continuous Lorentzian spectral density,
\begin{align}\label{eq:LSD}
	J_L(\omega) = \frac{\gamma_0}{2\pi} 
	\frac{\lambda^2}{\lambda^2 + \left(\omega - \omega_0 + \Delta\right)^2},
\end{align}
where $\lambda$ denotes the width of the Lorentzian, and $\Delta$ represents the detuning between the bare system 
frequency $\omega_0$ and the peak of the spectral distribution. The prefactor $\gamma_0$ in Eq.~\eqref{eq:LSD}
can be interpreted as spontaneous emission rate in the Born-Markov limit. This can be seen from 
Eq.~\eqref{gamma2-Lorentzian}
which gives $\gamma(t)$ to second order in the system-environment coupling. Taking the limit of an infinite spectral width 
$\lambda \to \infty$, one obtains $\gamma(t) \to \gamma_0$, showing that
$\gamma_0$ represents the Markovian decay rate. We note that $\gamma_0$ is of second order in the 
system-environment coupling as the spectral density is obviously a quantity of second order. 
Let us introduce a dimensionless expansion parameter $\alpha$ by means of the relation
\begin{align}\label{eq:Alpha}
	\alpha^2 = \frac{\gamma_0}{\lambda} = \frac{\tau_B}{\tau_R}.
\end{align}
Here, $\gamma_0$ is associated with the inverse of the system characteristic relaxation timescale $\tau_R$, 
while $\lambda$ corresponds to the inverse of the bath correlation time $\tau_B$, which can be seen from the
expression \eqref{eq:MemK} for the memory kernel below. In the limiting case of a 
rapidly evolving environment, $\tau_B \ll \tau_R$, the master equation (\Ref{eq:FAME}) reduces to the standard Lindblad 
form, corresponding to a purely Markovian description. In this regime, environmental processes occur on much shorter 
timescales than the system dynamics, and any information transferred from the system to the bath is effectively 
lost. The Lorentzian spectral density thus provides a convenient framework for studying both Markovian and non-Markovian 
regimes within the TCL formalism: by tuning the ratio $\alpha^2=\tau_B / \tau_R$ and the detuning $\Delta$, one can 
systematically explore how environmental memory effects influence the reduced system dynamics.

To ensure the infrared finiteness of frequency integrals over the spectral density, the Lorentzian distribution of
Eq.~\eqref{eq:LSD} is modified in the low-frequency range $\omega \leq \omega_m$.\cite{Irene} We introduce an exponent 
$\Omega$ that allows for an Ohmic ($\Omega = 1$) or super-Ohmic ($\Omega > 1$) low-frequency behavior,
\begin{align}\label{eq:mLSD}
	J(\omega) = \begin{cases} 
		\left(\frac{\omega}{\omega_m}\right)^{\Omega} J_L(\omega), & 0 \leq \omega \leq \omega_m, \\ 
		J_L(\omega), & \omega \geq \omega_m.
	\end{cases}
\end{align}
In our numerical simulations presented below we have used a super-Ohmic exponent $\Omega=2$.
For the calculation of the memory kernel we assume that contributions from this low-frequency modification are negligible and extend the integral over the entire real frequency axis,\cite{Irene}
\begin{align}\label{eq:MemK}
	\mathcal{K}(t) &= \int_0^\infty d\omega \, J(\omega) e^{-i\omega t} 
	\approx \int_{-\infty}^\infty d\omega \, J_L(\omega) e^{-i\omega t} \nonumber \\
	&= \frac{\gamma_0 \lambda}{2} \, e^{-\left(\lambda + i(\omega_0 - \Delta)\right)t}.
\end{align}
This yields an oscillating, exponentially decaying memory kernel, where the decay rate is set by the spectral linewidth 
$\lambda$.


\subsection{TCL Expansion}\label{TCL-Expansion-FA-Model}

As mentioned already we will carry out the TCL expansion directly by an expansion of the time-dependent coefficients 
$\gamma(t)$, $\gamma_{+}(t)$ and $\omega_{r}(t)$ 
given in Eqs.~\eqref{eq:CoeffExactwr}, \eqref{eq:CoeffExactGam} and \eqref{eq:CoeffExactGamP}. 
Accordingly, the perturbative expansion is realized through an expansion of the Green function $G(t)$ in powers of the 
coupling strength
\begin{align}\label{eq:Expansion2nd}
	G(t) &\approx G^{(0)}(t) + G^{(2)}(t) + G^{(4)}(t) + \dots ,
\end{align}
where the superscript $(n)$ denotes contributions of order $\alpha^n$ in the system-environment coupling.
In the present model, only even orders appear due to the choice of the initial state for the bath.
The free system evolution, given by the zeroth order contribution, follows from \eqref{eq:IDE} by taking into account that the 
memory kernel is of second order. The free evolution is thereby described by the Green function 
$G^{(0)}(t) = e^{-i\omega_0 t}$ with the bare system frequency $\omega_0$, solving the zeroth coupling order differential 
equation
\begin{align}\label{eq:DE0th}
	\dot{G}^{(0)}(t) + i\omega_0 G^{(0)}(t) = 0.
\end{align}

\subsubsection{Second Order}

To obtain the coefficients at the first non-vanishing order - that is, when the weak-coupling approximation is applied by 
retaining the leading perturbative corrections - we expand the Green function and collect the resulting terms according to 
their order in the coupling strength:
\begin{align}
	\gamma(t) &= -2\,\mathrm{Re}\!\left(\frac{\dot{G}(t)}{G(t)}\right)
	\approx -2\,\mathrm{Re}\!\left(
	\underbrace{\frac{\dot{G}^{(0)}(t)}{G^{(0)}(t)}}_{=-i\omega_0}
	+ \frac{d}{dt}\!\left(\frac{G^{(2)}(t)}{G^{(0)}(t)}\right)
	\right) \nonumber \\
	&= -2\,\mathrm{Re}\!\left(\dot{\tilde{G}}^{(2)}(t)\right)
	\equiv \gamma^{(2)}(t) . \label{gamma2(t)}
\end{align}
Here, we introduced the decomposition $G(t) = \tilde{G}(t) G^{(0)}(t)$, where $G^{(0)}(t)$ represents the free evolution of 
the system, and $\tilde{G}(t)$ captures the corrections due to system environment coupling. With this the frequency 
renormalization is analogously given by a time dependent frequency shift
\begin{align} \label{omega2(t)}
	\omega_r(t) &\approx \omega_0 - \mathrm{Im}\left(\dot{\tilde{G}}^{(2)}(t)\right) = \omega_0 + \omega_{r}^{(2)}(t).
\end{align}
Since both the spontaneous emission rate and the noise integral, which together determine the absorption rate, are of 
second order, they do not jointly contribute in the weak-coupling scenario. The lowest order expansion of the noise integral 
is nevertheless calculated to determine its derivative and thus the absorption rate:
\begin{align}
	\mathcal{I}^{(2)}(t) &= \int_0^{\infty} d\omega J(\omega)n(\omega) \left|\int_0^t d\tau 
	G^{(0)}(\tau)e^{i\omega\tau}\right|^2 \nonumber  \\
	&= \int_0^{\infty} d\omega J(\omega)n(\omega) 
	\frac{\sin^2\left(\frac{\omega-\omega_0}{2}t\right)}{\left(\frac{\omega-\omega_0}{2}\right)^2} , \\
	\gamma_{+}^{(2)}(t) &= \dot{\mathcal{I}}^{(2)}(t) = \int_0^\infty d\omega J(\omega)n(\omega) 
	2\frac{\sin\left((\omega-\omega_0)t\right)}{\omega-\omega_0} \label{eq:Idot2}.
\end{align}
The calculation of the reduced Green function is done by implementing the separation of the free evolution into
\begin{align}
	\dot{G}^{(2)}(t) + i\omega_0 G^{(2)}(t) + \int_0^t d\tau \mathcal{K}(t-\tau) G^{(0)}(\tau) = 0 ,
\end{align}
resulting in the expression required for calculating the spontaneous emission rate and frequency renormalization 
effects [Eqs.~\eqref{eq:CoeffExactwr}, \eqref{eq:CoeffExactGam} and \eqref{eq:CoeffExactGamP}]:
\begin{align} \label{dot-G-tilde}
	\dot{\tilde{G}}^{(2)}(t) &= -\int_0^t d\tau \mathcal{K}\left(t-\tau\right) e^{i\omega_0\left(t-\tau\right)} .
\end{align}


\subsubsection{Fourth Order}

In fourth order we get the contributions to the renormalized frequency,
\begin{align}
	\omega_r^{(4)}(t) &= -\mathrm{Im}\left( \dot{\tilde{G}}^{(4)}(t)-\dot{\tilde{G}}^{(2)}(t)\tilde{G}^{(2)}(t) \right), 
\end{align}
the spontaneous emission rate,
\begin{align}
	\gamma^{(4)}(t) &= -2\mathrm{Re}\left( \dot{\tilde{G}}^{(4)}(t)-\dot{\tilde{G}}^{(2)}(t)\tilde{G}^{(2)}(t) \right) ,
\end{align}
and to the noise integral
\begin{align}
	\mathcal{I}^{(4)}&(t) = \int_0^\infty d\omega 2J(\omega)n(\omega) \nonumber \\
	&\operatorname{Re} \left( \int_0^t d\tau G^{(2)}(\tau)e^{i\omega\tau} \int_0^t d\tau e^{-i(\omega-\omega_0)\tau} \right).
\end{align}
Altogether, they lead to the emission and absorption rates of the master equation given by:
\begin{align}
	\gamma_-(t) &= \gamma(t) + \gamma \mathcal{I}(t) + \dot{\mathcal{I}}(t) \\
	&\approx \gamma^{(2)}(t) + \gamma^{(4)}(t) + \gamma_+^{(2)}(t) + \gamma_+^{(4)}(t), \\
	\gamma_+(t) &\approx \underbrace{\dot{\mathcal{I}}^{(2)}(t)}_{\gamma_+^{(2)}(t)} + \underbrace{\dot{\mathcal{I}}^{(4)}(t) + \gamma^{(2)}(t) \mathcal{I}^{(2)}(t)}_{\gamma_+^{(4)}(t)}.
\end{align}
To find the Green function the differential equation \eqref{eq:IDE} has to be solved in fourth order:
\begin{align}\label{eq:Gdot4}
	\frac{d}{dt} G^{(4)}(t) + i\omega_0 G^{(4)}(t) + \int_0^t d\tau \mathcal{K}(t-\tau) G^{(2)}(\tau) = 0,
\end{align}
which yields
\begin{align}
	\dot{\tilde{G}}^{(4)}(t) &= -\int_0^t d\tau \mathcal{K}(\tau-t)e^{i\omega_0\left( t-\tau \right)}\tilde{G}^{(2)}(\tau) .
\end{align}


\section{Validity and Limitations of the TCL Expansion}\label{sec:Results}

In this section we compare the TCL expansion for a Lorentzian spectral density up to fourth order with the exact dynamics 
to assess convergence and accuracy of the expansion. Analytical and numerical results are presented, highlighting the 
perturbative regime and its dependence on coupling strength and detuning.\cite{TimMT}

\subsection{Exact Expressions}\label{sec:ExactOB}

In the case of a Lorentzian spectral density [Eq.~\eqref{eq:LSD}] it is possible to find an analytical result for the Green 
function, assuming that the approximation \eqref{eq:MemK} for the memory kernel holds:\cite{Irene, IreneMT}
\begin{align}\label{eq:exG}
	G\left(t\right) = \frac{e^{-i\omega_0t}}{s_2-s_1}\left(s_2e^{s_1t}-s_1e^{s_2t}\right),
\end{align}
where
\begin{align}
	s_{1,2} = \frac{1}{2}\left(-\lambda+i\Delta\pm\sqrt{\left(\lambda-i\Delta\right)^2-2\gamma_0\lambda}\right).
\end{align} 
To perform an expansion of this equation we rewrite the square root as
\begin{align}
	\left(\lambda-i\Delta\right)\sqrt{1-\frac{2\gamma_0\lambda}{\left(\lambda-i\Delta\right)^2}}.
\end{align}
An expansion in the parameter $\alpha^2=\frac{\gamma_0}{\lambda}$ is therefore possible as long as 
$\abs{\frac{2\gamma_0\lambda}{\left(\lambda-i\Delta\right)^2}}<1$ holds. This leads to a radius of convergence $R$ for
the expansion parameter $\alpha^2=\frac{\gamma_0}{\lambda}$ which is given by
\begin{align}\label{eq:RadiusOfC}
	R=\frac{1}{2}\left(1+\frac{\Delta^2}{\lambda^2}\right).
\end{align}
Beyond this radius of convergence, the perturbative expansion breaks down due to the non-analyticity of the square root 
at the branch point.
The parameter regimes can thus be classified into a weak-coupling regime, defined by $\gamma_0 / \lambda < R$, in 
which the perturbative expansion is valid, and a strong-coupling regime, $\gamma_0 / \lambda > R$, where the expansion 
no longer converges. For ratios $\gamma_0 / \lambda \leq 1/2$, the expansion is always valid, whereas for higher coupling 
strengths, the validity of the perturbative approach crucially depends on the detuning $\Delta$ between the system and the 
bath (see Sec.~\ref{sec:Numerics}).

Using the exact Green function \eqref{eq:exG} yields the exact expressions
\begin{align}
	&\omega_r\left(t\right) = \omega_0 - \mathrm{Im}\left( s_1s_2\frac{e^{s_1t}-e^{s_2t}}{s_2e^{s_1t}-s_1e^{s_2t}} \right) , \\
	&\gamma\left(t\right) = -2\mathrm{Re}\left( s_1s_2\frac{e^{s_1t}-e^{s_2t}}{s_2e^{s_1t}-s_1e^{s_2t}} \right) .
\end{align}
In the long term limit, in which the system should reach a stationary state and the coefficients of the master equation
become time independent, one obtains using $\mathrm{Re}\left(s_1\right)>\mathrm{Re}\left(s_2\right)$ the following 
expansion,
\begin{align}
	\omega_{r,st} &= \omega_0 - \mathrm{Im}\left(s_1\right) \nonumber \\
	&\approx \omega_0 + \frac{\gamma_0\lambda\Delta}{2\left(\lambda^2+\Delta^2\right)} + \frac{\gamma_0^2\lambda^2\Delta\left(\Delta^2-3\lambda^2\right)}{4\left(\lambda^2+\Delta^2\right)^3} , \\
	\gamma_{st} &\approx \frac{\gamma_0\lambda^2}{\lambda^2+\Delta^2} - \frac{\gamma_0^2\lambda^3\left(\lambda^2-3\Delta^2\right)}{2\left(\lambda^2+\Delta^2\right)^3},
\end{align}
where the index $st$ denotes the stationary value reached in the infinite time limit. 


\subsection{Second Order}

Using the memory kernel \eqref{eq:MemK} in Eq.~\eqref{dot-G-tilde} we obtain
\begin{align}
	\dot{\tilde{G}}^{(2)}(t) &= -\int_0^t d\tau \mathcal{K}(t - \tau) e^{i\omega_0 (t - \tau)} \nonumber \\
	&= \frac{\gamma_0\lambda}{2\left(-\lambda+i\Delta\right)} \left( 1 - e^{\left(-\lambda+i\Delta\right) t} \right).
\end{align}
Together with the initial condition $G(0)=1$ this yields:
\begin{align}\label{eq:G2}
	\tilde{G}^{(2)}(t) = \frac{\gamma_0\lambda}{2\left(-\lambda+i\Delta\right)}\left( \frac{1}{-\lambda+i\Delta} + t - \frac{e^{\left(-\lambda+i\Delta\right)t}}{-\lambda+i\Delta} \right) .
\end{align}
With the help of Eqs.~\eqref{gamma2(t)} and \eqref{omega2(t)} we get the time evolution of renormalized frequency and 
spontaneous emission rate:
\begin{align}
	\omega_r^{(2)}(t) &= -\mathrm{Im}\left( \dot{\tilde{G}}^{(2)}(t) \right) = \frac{\gamma_0 \lambda \Delta}{2(\lambda^2 + 
	\Delta^2)} \left( 1 - \cos(\Delta t) e^{-\lambda t} \right) \nonumber \\
	&- \frac{\gamma_0 \lambda^2}{2(\lambda^2 + \Delta^2)} \sin(\Delta t) e^{-\lambda t} , \\
	\gamma^{(2)}(t) &= -2 \operatorname{Re} \left( \dot{\tilde{G}}^{(2)}(t) \right) = \frac{\gamma_0 \lambda \Delta}
	{\lambda^2 + \Delta^2} \sin(\Delta t) e^{-\lambda t} \nonumber \\
	&+ \frac{\gamma_0 \lambda^2}{\lambda^2 + \Delta^2} \left( 1 - \cos(\Delta t) e^{-\lambda t} \right) . 
	\label{gamma2-Lorentzian}
\end{align}
Finding an analytical solution for the time evolution of the absorption rate is not possible due to the complexity of the noise 
integral \eqref{eq:Idot2}. However, in the long time limit a solution can be found, by extending the integration limit and 
using the representation of a $sinc$ function by a delta distribution.

To calculate the second order parameter $\gamma_{+}$ the time derivative of the noise integral \eqref{eq:NoiseInt}
has to be determined within second order, keeping in mind that the memory kernel is of second order in the coupling,
such that the Green function $G(t)$ has to be replaced by the zeroth order contribution $G^{(0)}(t) = e^{-i\omega_0 t}$.
This yields
\begin{align}\label{eq:NoiseInt2nd}
	\mathcal{I}^{(2)}(t) &= \int_0^\infty \! d\omega \, J(\omega)\, n(\omega)\,
	\left|\int_0^t \! d\tau \, e^{i\left(\omega-\omega_0\right)\tau}\right|^2 \nonumber \\
	&= \int_0^\infty \! d\omega \, J(\omega)\, n(\omega)\,
	\left( \frac{\sin\left(\frac{\omega-\omega_0}{2}t\right)}{\frac{\omega-\omega_0}{2}} \right)^2,
\end{align}
and the time derivative is given by
\begin{align}
	\dot{\mathcal{I}}^{(2)}(t) = \int_0^\infty \! d\omega \, J(\omega)\, n(\omega)\,
	2\frac{\sin\left[\left(\omega-\omega_0\right)t\right]}{\omega-\omega_0}.
\end{align}
In the limit $t\to\infty$ the term $\frac{\sin\left(\left(\omega-\omega_0\right)t\right)}{\pi\left(\omega-\omega_0\right)}$ 
approaches the delta function $\delta(\omega-\omega_0)$. Thus, in the steady state we find the second order
absorption rate
\begin{align}\label{eq:Idot2LSD}
	\gamma_{+,st}^{(2)} = \dot{\mathcal{I}}^{(2)}_{st} = 2\pi J(\omega_0) n(\omega_0) = \frac{\gamma_0 \lambda^2}{\lambda^2 + \Delta^2} n(\omega_0). \\ \nonumber
\end{align}


\subsection{Fourth Order}

To determine the fourth order we start from
\begin{eqnarray}
 \dot{\tilde{G}}^{(4)}(t) &=& -\int_0^t d\tau K(\tau-t)e^{i\omega_0\left( t-\tau \right)}\tilde{G}^{(2)}(\tau) \nonumber \\
 &=& \frac{\lambda^2\gamma_0^2}{4}\frac{\lambda + i\Delta}{\left(\lambda^2+\Delta^2\right)^2}  \\
 && \left[ t + te^{\left(-\lambda+i\Delta\right)t}
 - \frac{\lambda + i\Delta}{\lambda^2+\Delta^2}\left( 1-e^{\left(-\lambda+i\Delta\right)t} \right)  \right] . \nonumber
\end{eqnarray}
This is the used to determine the spontaneous emission rate and the frequency renormalization in fourth order:
\begin{widetext}
	\begin{align}
		\gamma^{(4)}(t) &= -2\mathrm{Re}\left( \dot{\tilde{G}}^{(4)}(t)-\dot{\tilde{G}}^{(2)}(t)\tilde{G}^{(2)}(t) \right) \nonumber \\
		&= \frac{\gamma_0^2\lambda^5 e^{-\lambda t}}{2\left(\lambda^2+\Delta^2\right)^3}\left\{ \left(e^{\lambda t}-e^{-\lambda t}\cos\left(2\Delta t\right)\right)\left(1-3\frac{\Delta^2}{\lambda^2}\right)-2\lambda t\cos\left(\Delta t\right)\left(1-\frac{\Delta^4}{\lambda^4}\right) \right. \nonumber \\
		&\phantom{=}\left. +4\Delta t\left(1+\frac{\Delta^2}{\lambda^2}\right)\sin\left(\Delta t\right)+\frac{\Delta}{\lambda}e^{-\lambda t}\sin\left(2\Delta t\right)\left(3-\frac{\Delta^2}{\lambda^2}\right)\right\} ,
	\end{align}
	\begin{align}
		\omega_r^{(4)}(t) &= -\mathrm{Im}\left( \dot{\tilde{G}}^{(4)}(t)-\dot{\tilde{G}}^{(2)}(t)\tilde{G}^{(2)}(t) \right) \nonumber \\
		&= -\frac{\gamma_0^2\lambda^2 \Delta^3e^{-\lambda t}}{4\left(\lambda^2+\Delta^2\right)^3}\left\{ \left(e^{\lambda t}-e^{-\lambda t}\cos\left(2\Delta t\right)\right)\left(1-3\frac{\lambda^2}{\Delta^2}\right)-2\Delta t\sin\left(\Delta t\right)\left(1-\frac{\lambda^4}{\Delta^4}\right) \right. \nonumber \\
		&\phantom{=}\left. +4\lambda t\left(1+\frac{\lambda^2}{\Delta^2}\right)\cos\left(\Delta t\right)-\frac{\lambda}{\Delta}e^{-\lambda t}\sin\left(2\Delta t\right)\left(3-\frac{\lambda^2}{\Delta^2}\right)\right\} .
	\end{align}
\end{widetext}
The corresponding stationary fourth order contributions are easily obtained by taking the long time limit, yielding
\begin{align}
	\gamma^{(4)}_{st} &= \frac{\gamma_0^2\lambda^5 }{2\left(\lambda^2+\Delta^2\right)^3}\left(1-3\frac{\Delta^2}{\lambda^2}\right) , \label{eq:gamma4} \\
	\omega^{(4)}_{r,st} &= \frac{\gamma_0^2\lambda^2 \Delta^3}{4\left(\lambda^2+\Delta^2\right)^3}\left(3\frac{\lambda^2}{\Delta^2}-1\right) . \label{eq:w_r4}
\end{align}
The remaining contribution is the absorption rate $\gamma_{+}(t)$, where the time dependent values cannot 
be calculated analytically. We start from
\begin{align}
	\mathcal{I}^{(4)}&(t) = \int_0^\infty d\omega 2J(\omega)n(\omega) \nonumber \\
	&\operatorname{Re} \left( \int_0^t d\tau G^{(2)}(\tau)e^{i\omega\tau} \int_0^t d\tau e^{-i(\omega-\omega_0)\tau} \right),
\end{align}
and calculate the time derivative:
\begin{align}\label{eq:APdotI4}
	\dot{\mathcal{I}}^{(4)}(t) &= \int_0^\infty d\omega J(\omega)n(\omega) \nonumber \\
	& 2 \operatorname{Re} \left( \int_0^t d\tau \tilde{G}^{(2)}(\tau)e^{i(\omega-\omega_0)\left(\tau-t\right)} \right. 
	\nonumber \\
	& \quad\quad \left. + \tilde{G}^{(2)}(t)e^{i(\omega-\omega_0)t} \int_0^t d\tau e^{-i(\omega-\omega_0)\tau} \right).
\end{align}
This expression can be calculated again within the steady state limit, replacing the sinc function
by a delta function. This leads to:
\begin{align}
	&\dot{\mathcal{I}}^{(4)}_{st} = -2J(\omega_0)n(\omega_0)t\pi\frac{\gamma_0\lambda^2}{\lambda^2+\Delta^2} \nonumber \\
	& +2\pi J(\omega_0) n(\omega_0) \lambda \gamma_0 
	\frac{\lambda^2 - \Delta^2}{(\lambda^2 + \Delta^2)^2} \nonumber \\
	&- \lambda \gamma_0 \Delta 
	\frac{\pi n(\omega_0) J(\omega_0)}{\lambda^2 + \Delta^2}
	\left(
	e^{\omega_0 \beta}n(\omega_0)\beta
	+ \frac{2\Delta}{\lambda^2 + \Delta^2}
	\right) .
\end{align}
When calculating the steady state absorption rate, note that the term linearly increasing with time $t$ in this expression is 
cancelled against a term in $G^{(2)}(t)$ which also increases linearly with $t$ (see Eq.~\eqref{eq:G2}). Thus, the steady 
state absorption rate becomes
\begin{align} \label{eq:gamP4Os}
	\gamma_{+,st}^{(4)} =& \gamma_{st}^{(2)}\mathcal{I}_{st}^{(2)} + \dot{\mathcal{I}}_{st}^{(4)} \nonumber \\
	=& 2\pi J(\omega_0) n(\omega_0) \lambda \gamma_0
	\frac{\lambda^2 - \Delta^2}{(\lambda^2 + \Delta^2)^2} \nonumber \\
	&- \lambda \gamma_0 \Delta 
	\frac{\pi n(\omega_0) J(\omega_0)}{\lambda^2 + \Delta^2}
	\left(
	e^{\omega_0 \beta}n(\omega_0)\beta
	+ \frac{2\Delta}{\lambda^2 + \Delta^2}
	\right).
\end{align}
Here the importance of modifying the Lorentzian spectral density is evident, since the initial environmental Planck 
distribution $n(\omega)$ diverges for $\omega\to 0$, the product with the spectral density $J(\omega)$ 
should be finite. Using the super-ohmic modification, the last term vanishes, while an ohmic spectral density
leads to an oscillating behavior. 

\begin{figure}[htb]
	\centering
	\includegraphics[width=0.9\columnwidth]{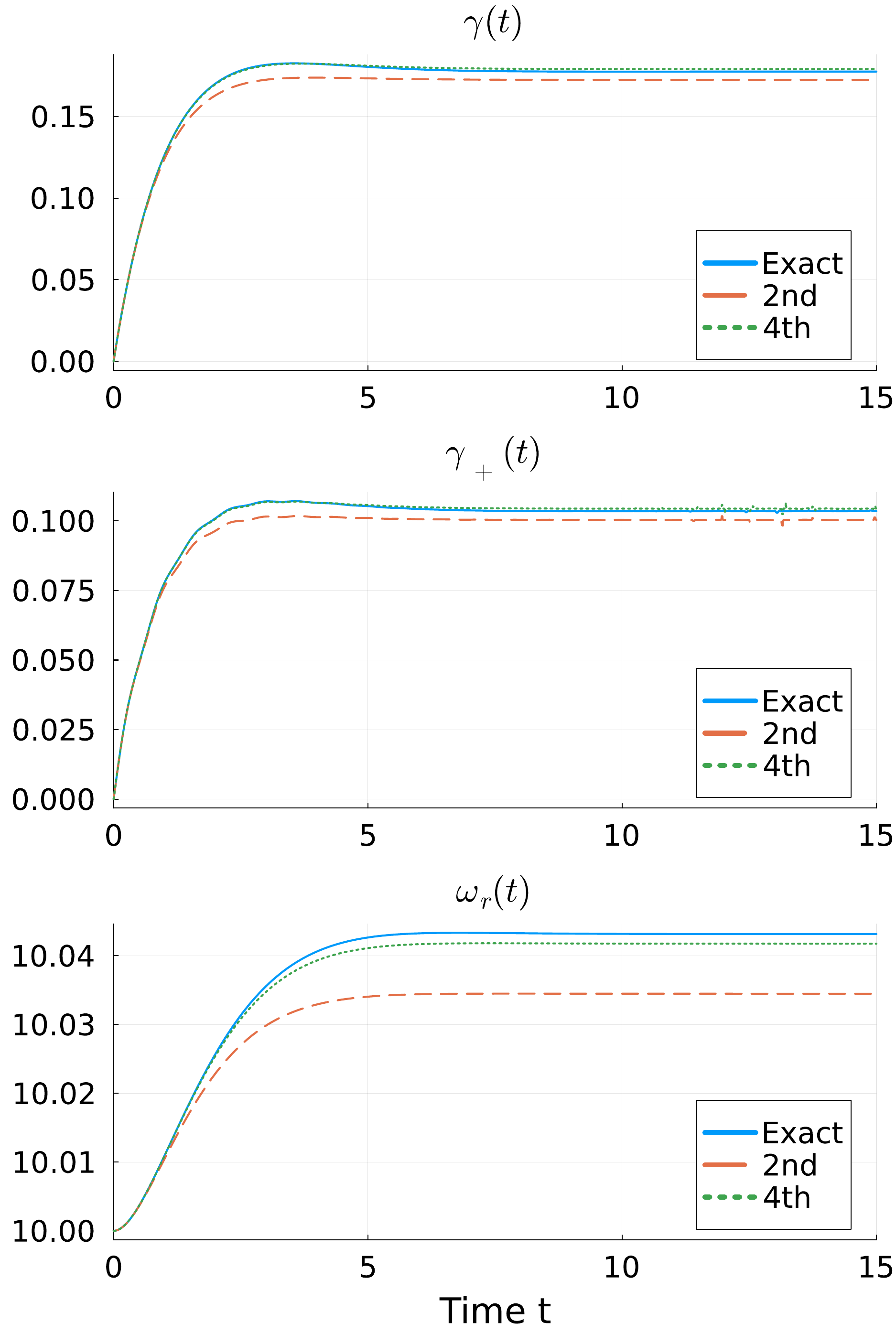}
	\caption{Time evolution of the coefficients of the master equation \eqref{eq:FAME} for a weak coupling strength 
	$\gamma_0 = 0.2$, with $\lambda = 1.0$ and $\Delta = 0.4$. The fourth-order expansion (green) closely matches the 
	exact result (blue), while the second-order expansion (orange) shows small deviations.}
	\label{fig:gP_weak}
\end{figure}

\subsection{Numerical Comparison} \label{sec:Numerics}

All quantities and parameters presented in the following numerical discussion are either dimensionless or have dimensions of energy, since we set $\hbar = k_B = 1$. Accordingly, all dimensionful quantities are expressed in terms of an arbitrary energy or frequency unit. The chosen parameter values for the numerical calculations are $\omega_0 = 10.0$, $T = 10.0$, and $\lambda = 1.0$. Here, $\lambda$ defines the width of the Lorentzian spectral density and determines the sharpness of the spectral peak around the central frequency. A smaller $\lambda$ corresponds to a narrowly peaked spectrum, in which the system interacts predominantly with bath modes near resonance, whereas a larger $\lambda$ produces a broader spectrum, allowing a wider range of bath frequencies to contribute. The chosen value $\lambda = 1.0$ represents a moderately broad peak. The system frequency $\omega_0$ and the temperature $T$ are selected such that thermal fluctuations and coherent dynamics occur on comparable energy scales, ensuring a nontrivial interplay between the system and its environment.

Regarding $\gamma_0$, we distinguish between weak and strong coupling regimes for the fixed parameters $\lambda=1.0$ and $\Delta=0.4$. For this setup, the radius of convergence $R$ (Eq.~\eqref{eq:RadiusOfC}) is given by $R(\Delta=0.4) = 0.58$, which, with the chosen value of $\lambda$, translates to an upper limit of $\gamma_0 = 0.58$ for the validity of the perturbative expansion. The corresponding time evolutions of the master equation parameters in a weak-coupling scenario with $\gamma_0 = 0.2$ are shown in~\Cref{fig:gP_weak}.

\begin{figure}[htb]
	\centering
	\includegraphics[width=0.9\columnwidth]{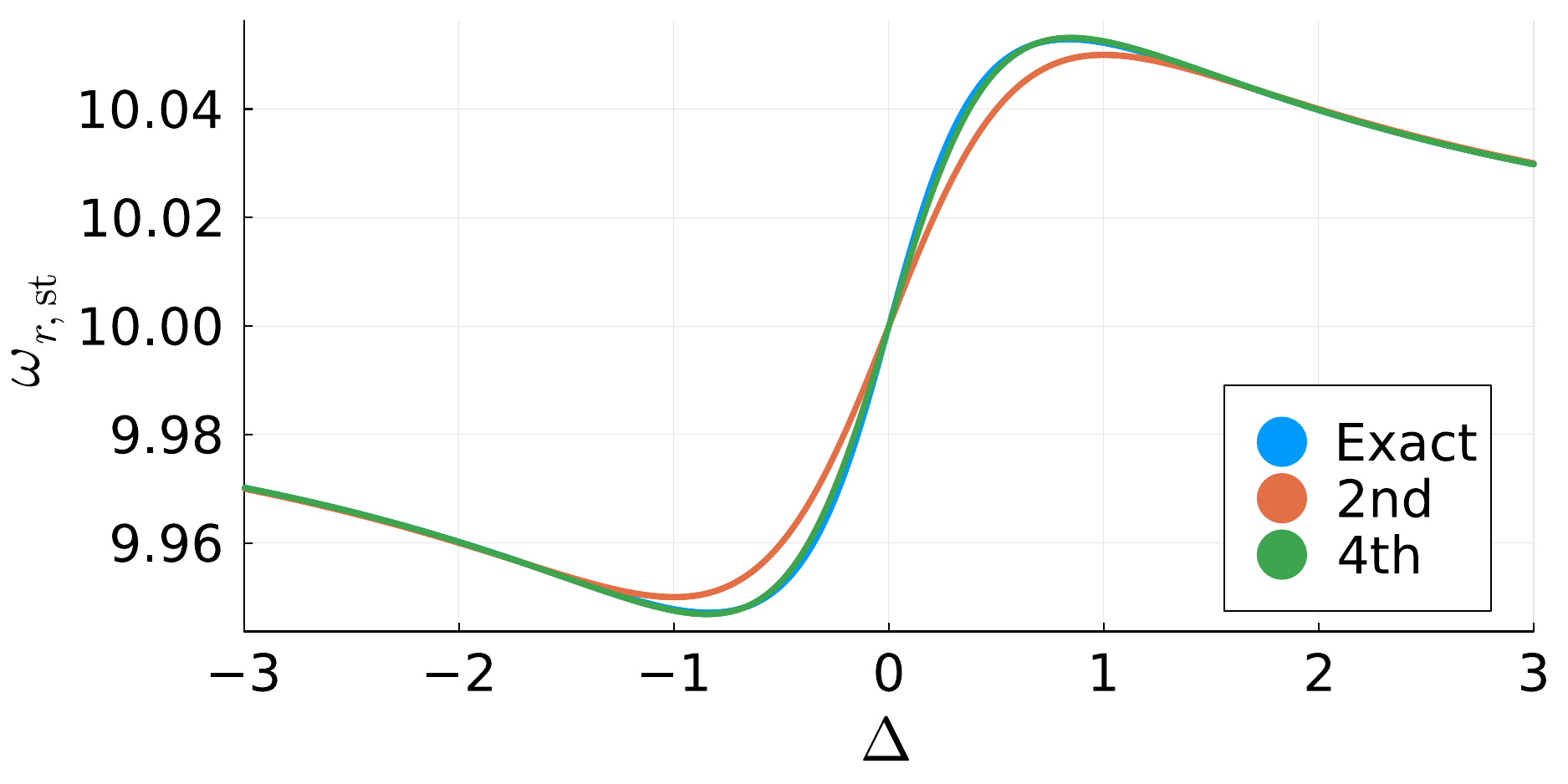}
	\caption{Steady-state values of $\omega_r$ as a function of the detuning $\Delta$, with weak coupling 
	$\gamma_0=0.2$ and $\lambda = 1.0$. The fourth-order expansion closely matches the exact results, while the 
	second-order expansion shows slight deviations. Since $\gamma_0 / \lambda < 1/2$, the perturbative treatment is 
	valid over the entire range of detunings.}
	\label{fig:wr_Delta}
\end{figure}

We observe that the accuracy of the time-dependent parameters improves with increasing perturbative order, closely reproducing the exact dynamics. For the weak-coupling case, the steady-state values of the renormalized frequency for a fixed $\gamma_0$ are shown in \Cref{fig:wr_Delta} in dependency on the detuning $\Delta$, demonstrating that the expansion approach also accurately captures the long-time behavior of the system. Including the fourth-order approximation further enhances the agreement with the exact dynamics, significantly improving the precision of the perturbative description.

\begin{figure}[htb]
	\centering
	\includegraphics[width=0.9\columnwidth]{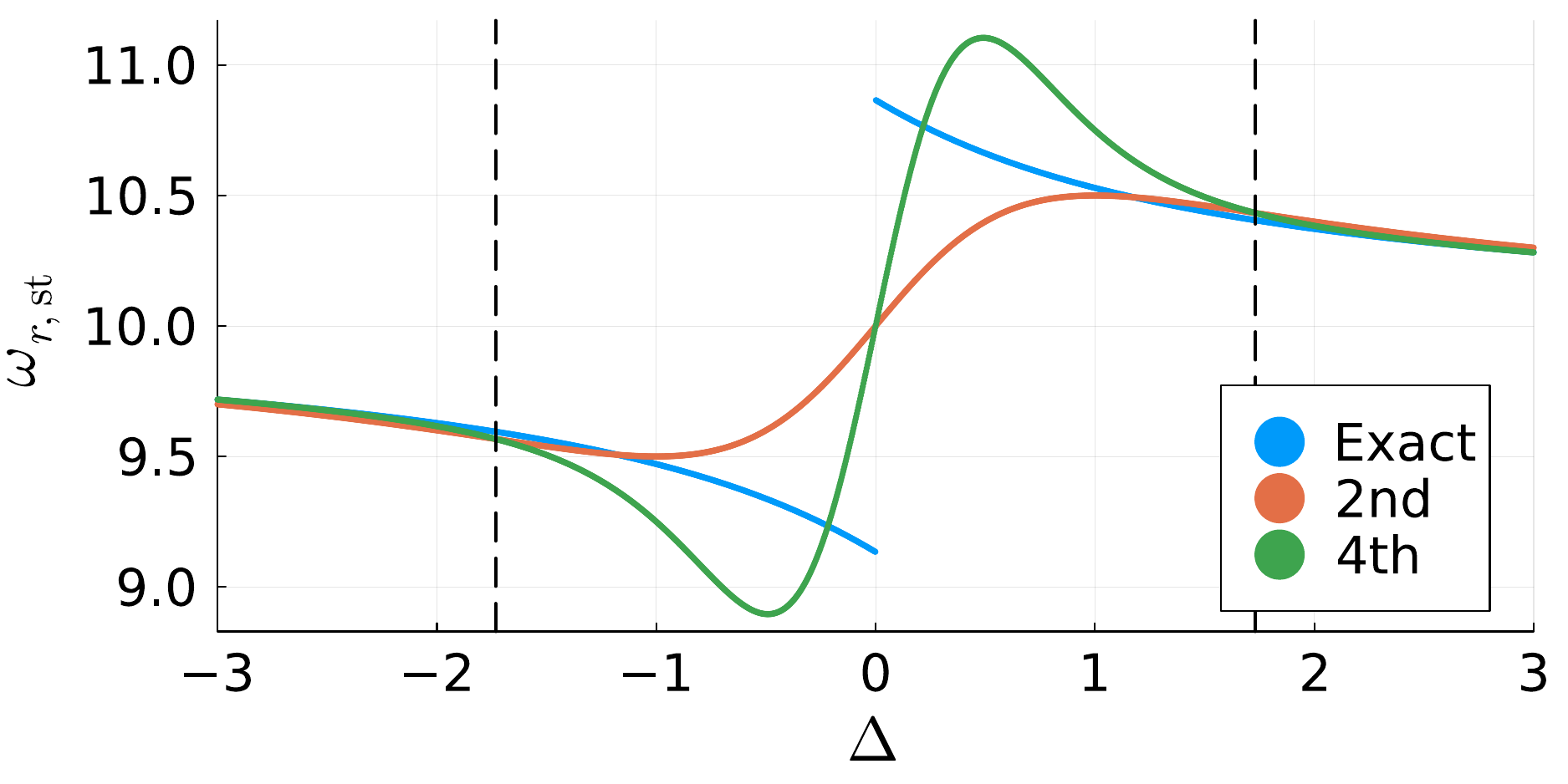}
	\caption{Steady-state values of $\omega_r$ for $\gamma_0 / \lambda = 2.0$. The dashed line indicates the detuning $\Delta$ at which the radius of convergence is reached, i.e., $R(\Delta) = \frac{1}{2}\left(1 + \frac{\Delta^2}{\lambda^2}\right) = 2.0$. Between these lines, the perturbative treatment fails, and a discontinuity appears in the exact solution at resonance. Outside this non-perturbative regime, the expansion accurately reproduces the exact steady-state values. }
	\label{fig:wr_Delta_st}
\end{figure}

In the strong-coupling regime, with $\gamma_0 = 2.0$, the validity of the perturbative expansion depends on the detuning parameter $\Delta$. The behavior of the steady-state renormalized frequency for different detunings is shown in~\Cref{fig:wr_Delta_st}. The TCL expansion fails to reproduce the steady-state values correctly when the detuning is sufficiently small such that $\gamma_0 / \lambda > R$. Specifically, the expansion remains valid only for $\lvert \Delta \rvert > \sqrt{\lambda (2 \gamma_0 - \lambda)}$, ensuring that the radius of convergence exceeds the expansion parameter. The discontinuity observed in the exact steady-state renormalized frequency at resonance occurs only in the strong-coupling regime, $\gamma_0 / \lambda > 1/2$, where the perturbative treatment breaks down. This discontinuity in the
strong coupling regime can be shown to be\cite{IreneMT}
\begin{equation}
 \omega_{r,\mathrm{st}}(\Delta \rightarrow +0) -  \omega_{r,\mathrm{st}}(\Delta \rightarrow -0)  
 = \sqrt{\lambda \left(2\gamma_0 - \lambda\right)}.
\end{equation}
In this context we remark that it is of great interest to study in more detail the influence of the
rotating wave approximation used in the microscopic model Hamiltonian \eqref{eq:H_SE} on the features of this
behavior of the renormalized frequency at resonance.

\begin{figure}[htb]
	\centering
	\includegraphics[width=0.9\columnwidth]{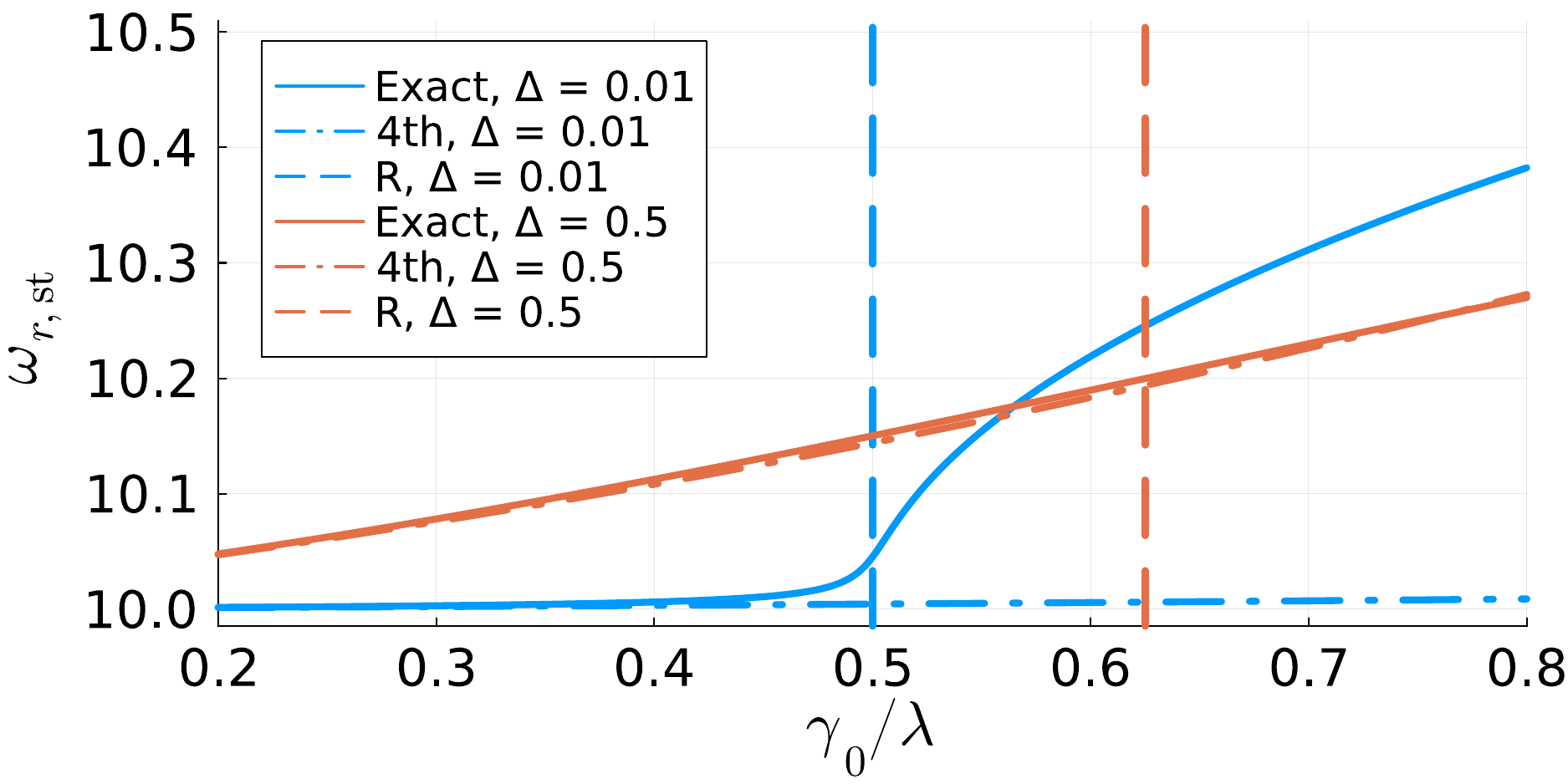}
	\caption{Stationary value of $\omega_{r}$ for varying detuning and coupling strenghts, calculated exactly and 
	perturbatively in fourth order. The closer to resonance the larger is the distinction between weak and strong coupling 
	regime, separated by the radius of convergent indicated by the dashed lines.}
	\label{fig:wr_Res}
\end{figure}

The dependence of the frequency renormalization on the coupling strength is depicted in \Cref{fig:wr_Res}. For different 
values of the detuning $\Delta$ the exact steady state values are shown together with the fourth order approximation. 
Additionally the radii of convergence are plotted. These radii not only separate different regimes in a mathematical sense, 
but also indicate a qualitative change in the system's physical behavior, highlighting the strong sensitivity of the system 
properties in this regime.

\begin{figure}[htb]
	\centering
	\includegraphics[width=0.9\columnwidth]{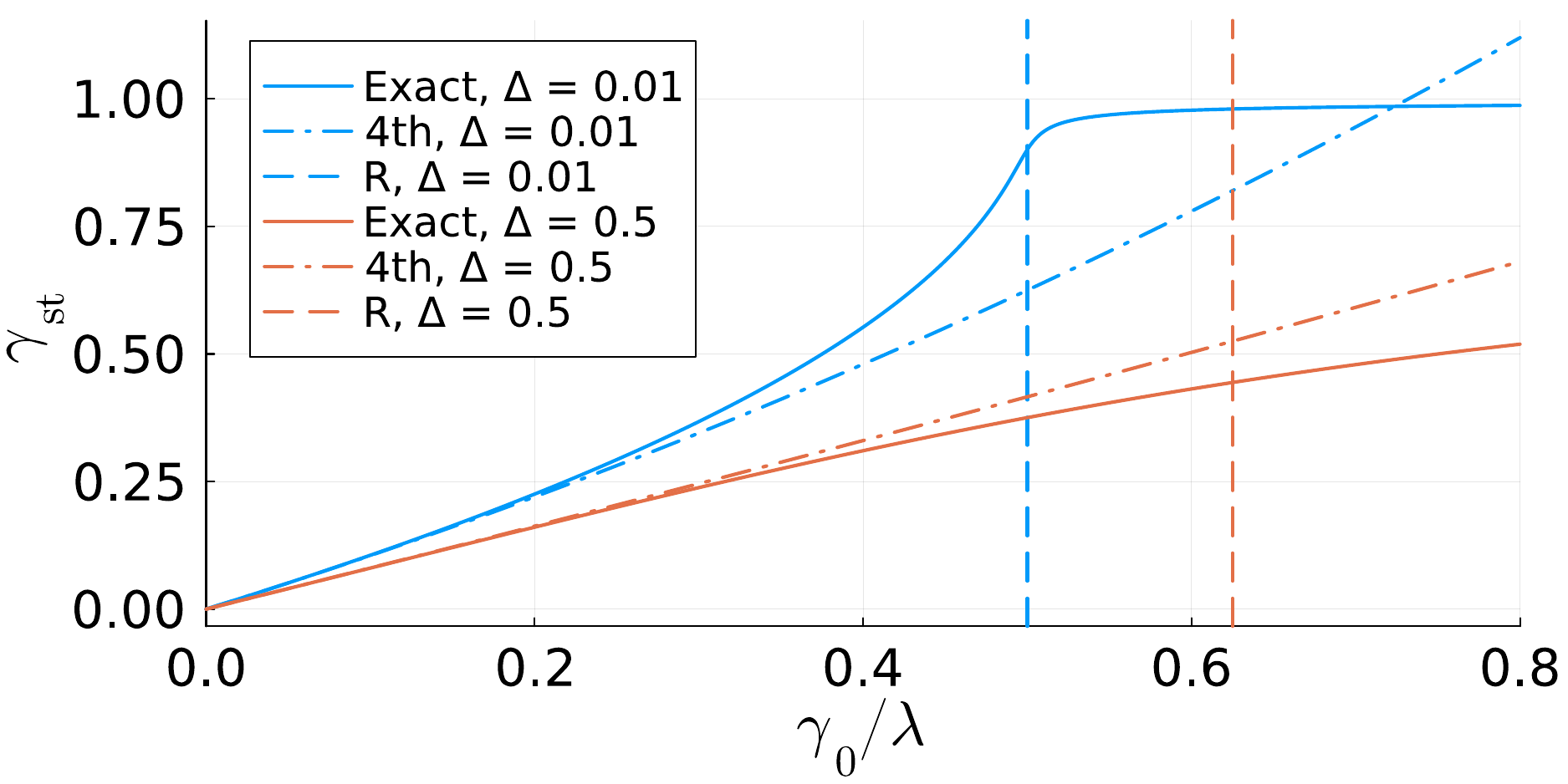}
	\caption{Steady-state values of $\gamma$ for varying detuning and coupling strengths, calculated exactly and using 
	the fourth-order expansion. Analogous to \Cref{fig:wr_Res}, the radius of convergence (dashed lines) separates weak 
	and strong coupling regime.}
	\label{fig:g_Res}
\end{figure}

Examining the spontaneous emission rate, shown in \Cref{fig:g_Res}, we observe a similar distinction between weak and strong coupling regimes near resonance, separated by the radius of convergence. The maximal steady-state rate occurs at resonance, while increasing the coupling strength beyond $\lambda/2$ does not further increase the steady-state value, which remains $\gamma_{\mathrm{st}} = \lambda$. 

\section{Non-Markovianity} \label{sec:NM}

Building on the previous results, this section investigates to which extend the TCL expansion is able to capture 
non-Markovian effects which are defined by the Bures distance \eqref{Bures-dist} and the non-Markovianity measure 
\eqref{eq:NM}. As the Bures distance for Gaussian states is completely given in terms of the first and second moments
we first determine the equations of motion for these moments directly from the 
master equation \eqref{eq:FAME} which yields:\cite{AleReservoirs}
\begin{align}
	&\frac{d}{dt} \exval{a}_t = -\left(i\omega_r(t)+\frac{\gamma(t)}{2}\right) \exval{a}_t , \label{eq:ODEa} \\
	&\frac{d}{dt} \exval{aa}_t =  -\left(2i\omega_r(t)+\gamma(t)\right) \exval{aa}_t , \label{eq:ODEaa}\\
	&\frac{d}{dt} \exval{a^\dagger a}_t = -\gamma(t) \exval{a^\dagger a}_t
	+ \gamma(t) \mathcal{I}(t) + \dot{\mathcal{I}}(t) \label{eq:ODEn}.
\end{align}
The exact dynamics of the Fano-Anderson model is given by the solution of these equations:
\begin{align}
	&\langle a \rangle_t = G(t)\langle a \rangle_0 , \label{eq:EoMExpVala} \\
	&\langle a a \rangle_t = G^2(t)\langle a a \rangle_0 , \label{eq:EoMExpValaa}\\
	&\langle a^\dagger a \rangle_t = |G(t)|^2 \langle a^\dagger a \rangle_0 + \mathcal{I}(t) , \label{eq:EoMExpValada}
\end{align}
The perturbative motion of the moments is obtained by solving the equations of motion
\eqref{eq:ODEa}-\eqref{eq:ODEn} using the second and fourth order approximation for the time dependent coefficients.
As initial states of the oscillator mode $a$ we take two different coherent states (displaced vacuum states) in the following.

\begin{figure}[htb]
	\centering
	\includegraphics[width=0.9\columnwidth]{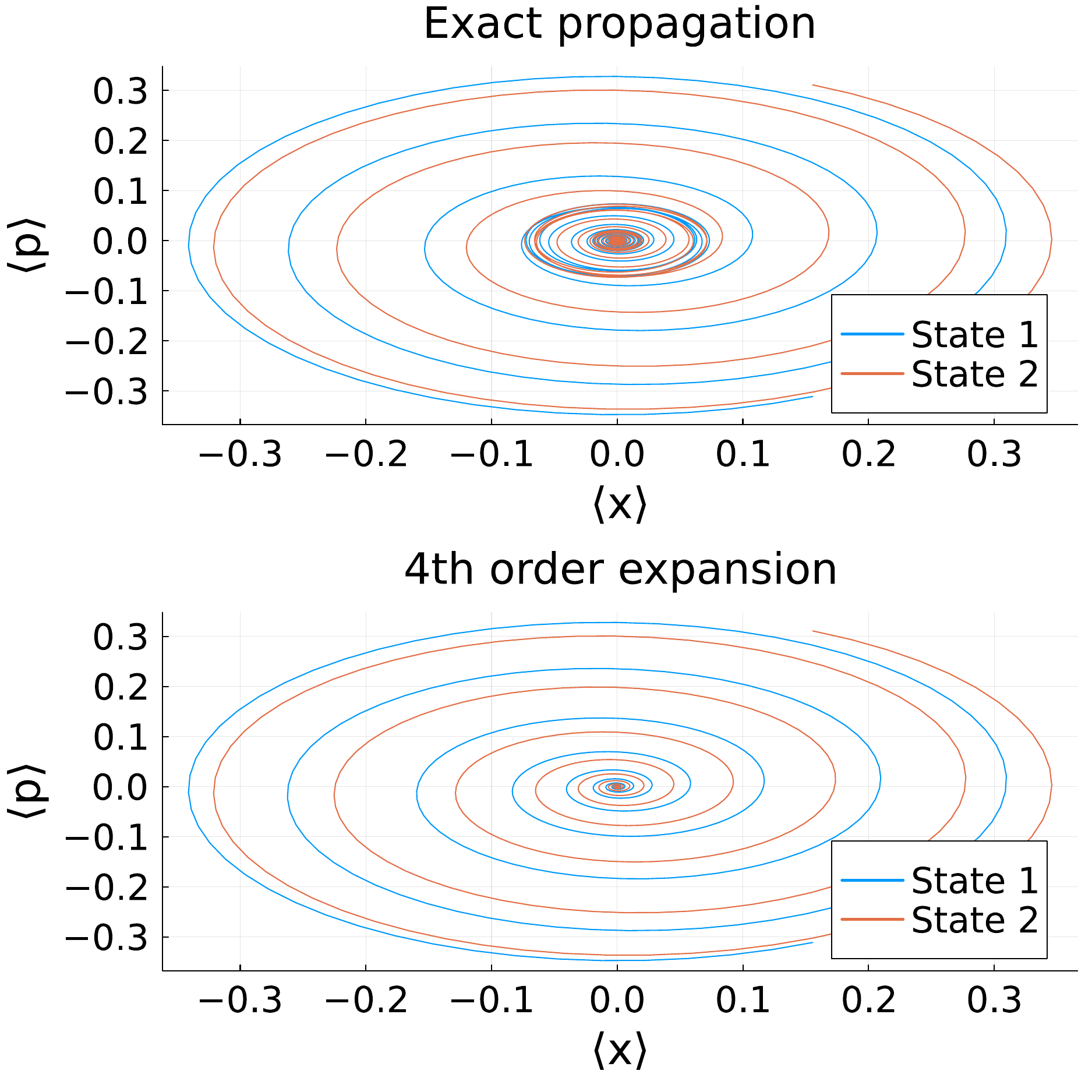}
	\caption{Phase-space evolution of two initial coherent states $|\alpha = 0.11 \pm 0.22i\rangle$, 
	computed both exactly and perturbatively for 
	$\gamma_0 = 2.0$ and $\Delta = 0.4$. While the exact propagation exhibits more complex trajectories, including temporary increases in distance from the states, the fourth-order perturbative expansion shows only a monotonic approach, failing to capture information backflow.}
	\label{fig:Phase_space_strong}
\end{figure}

First we show the phase space evolution of two initial coherent states in \Cref{fig:Phase_space_strong}. We use 
a strong coupling of $\gamma_0=2.0$ in this case in order to indicate that the exact motion can show crossings, 
approaching and diverging of the two evolutions, while in the approximate case both states approach the steady state 
monotonically. 

\begin{figure}[htb]
	\centering
	\includegraphics[width=0.9\columnwidth]{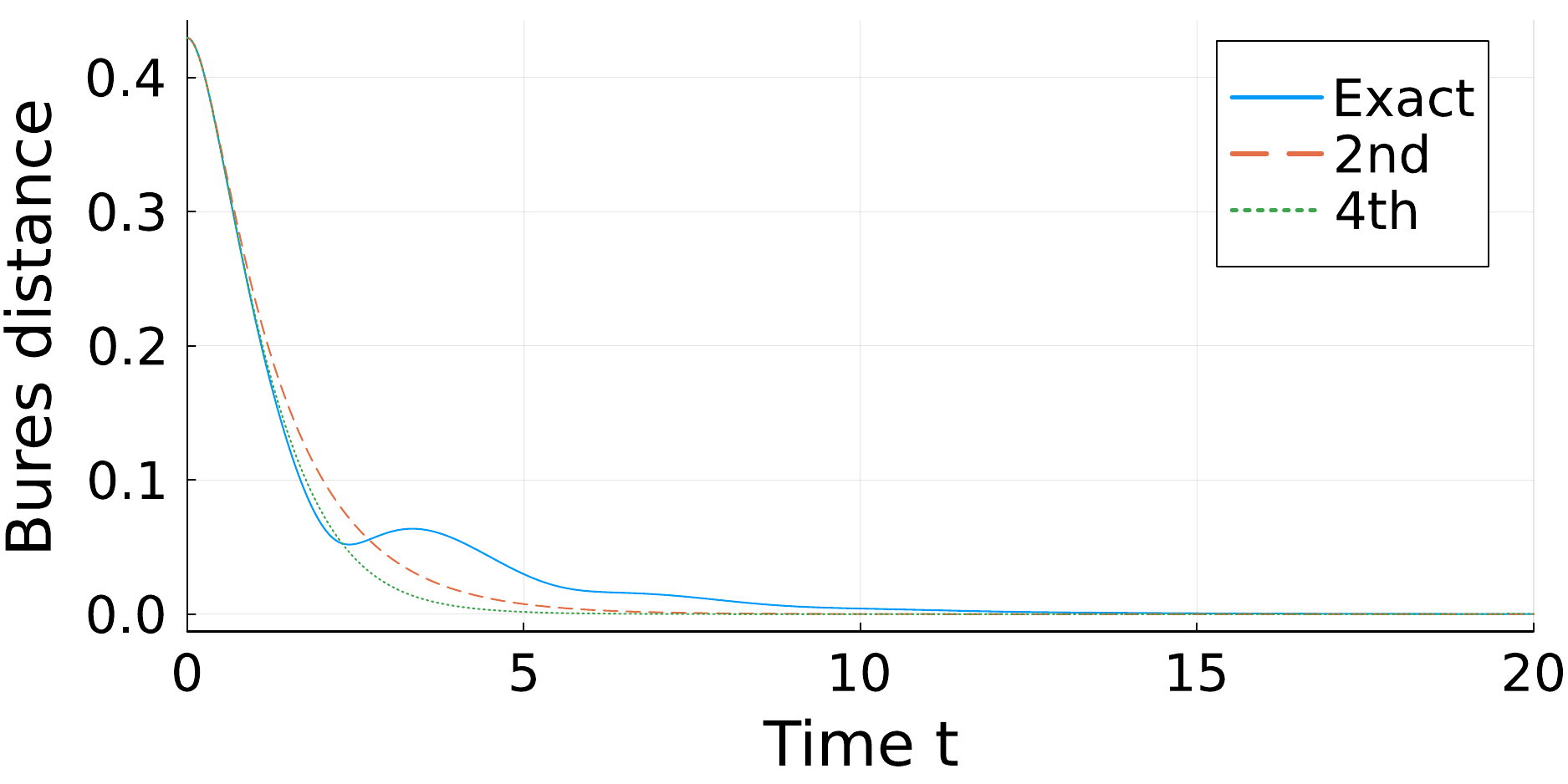}
	\caption{Bures distance for $\gamma_0 = 2.0$ and $\Delta = 0.4$. The coupling exceeds the perturbative convergence radius, so while the exact dynamics show revivals, the perturbative approach fails to capture them.}
	\label{fig:Bures_strong}
\end{figure}

\begin{figure}[htb]
	\centering
	\includegraphics[width=0.9\columnwidth]{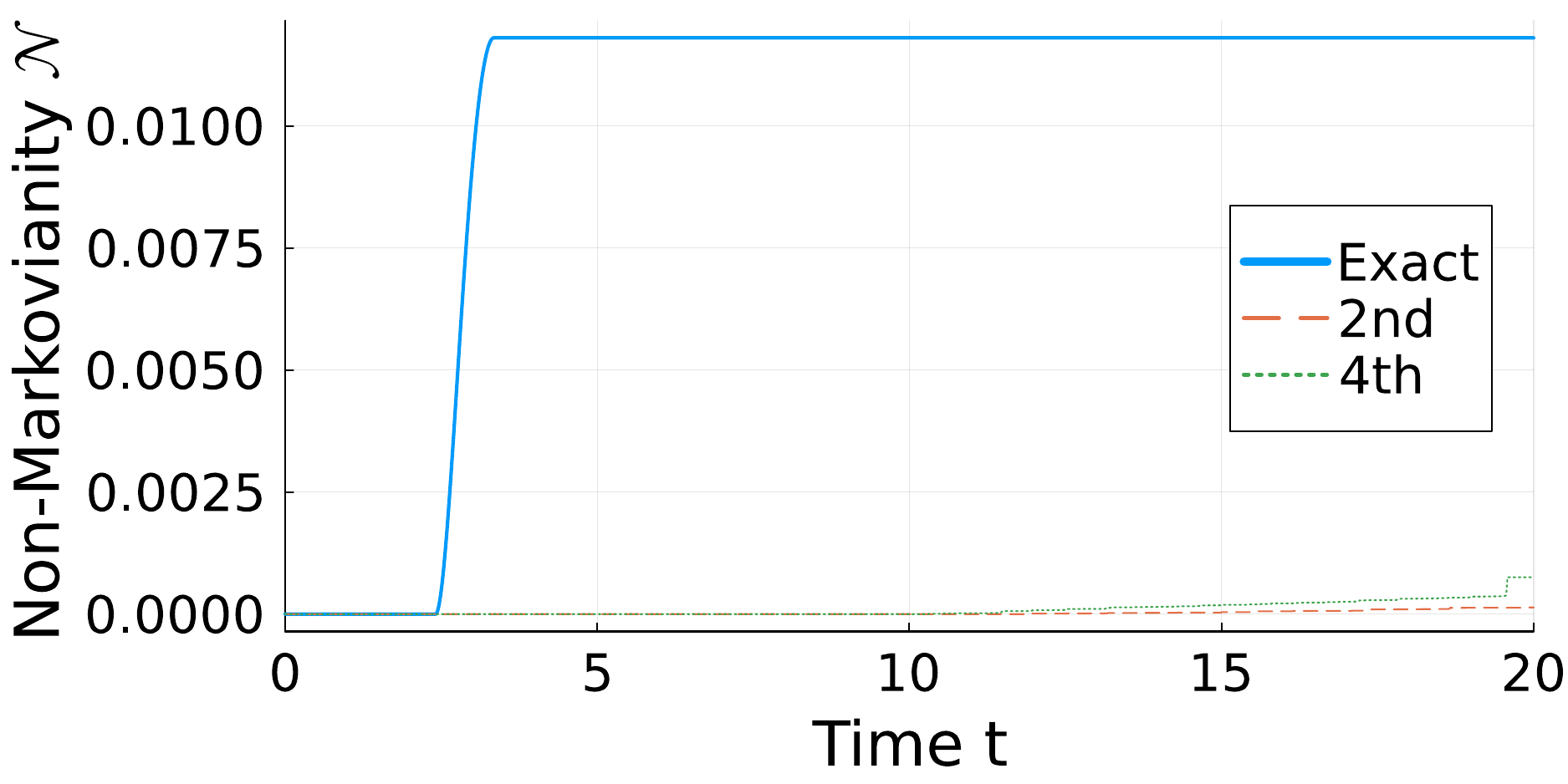}
	\caption{Measure of non-Markovianity for $\gamma_0 = 2.0$ and $\Delta = 0.4$, computed both exactly and perturbatively. The Bures-distance based measure $\mathcal{N}$ shows that the perturbative approach fails to capture information backflow. The slight linear increase in the perturbative curve is attributed to numerical inaccuracies.}
	\label{fig:NM_strong}
\end{figure}

When we transfer this to the Bures distance, displayed in \Cref{fig:Bures_strong}, we see that while the exact case contains increasing of the distance, the perturbative treatment is monotonically decreasing.
Since the perturbative expansion cannot capture this oscillations, and given that we define non-Markovianity~$\mathcal{N}$ 
as the total increase in Bures distance (see Eq.~\eqref{eq:NM}) this limitation manifests as a failure to represent
non-Markovian behavior in \Cref{fig:NM_strong}.
We attribute the tiny increase in non-Markovianity observed in the second- and fourth-order evolutions to numerical 
inaccuracies in the evaluation of the Bures distance. The approximately linear trend suggests that this effect results from 
small oscillations rather than genuine memory effects.

\begin{figure}[htb]
	\centering
	\includegraphics[width=0.9\columnwidth]{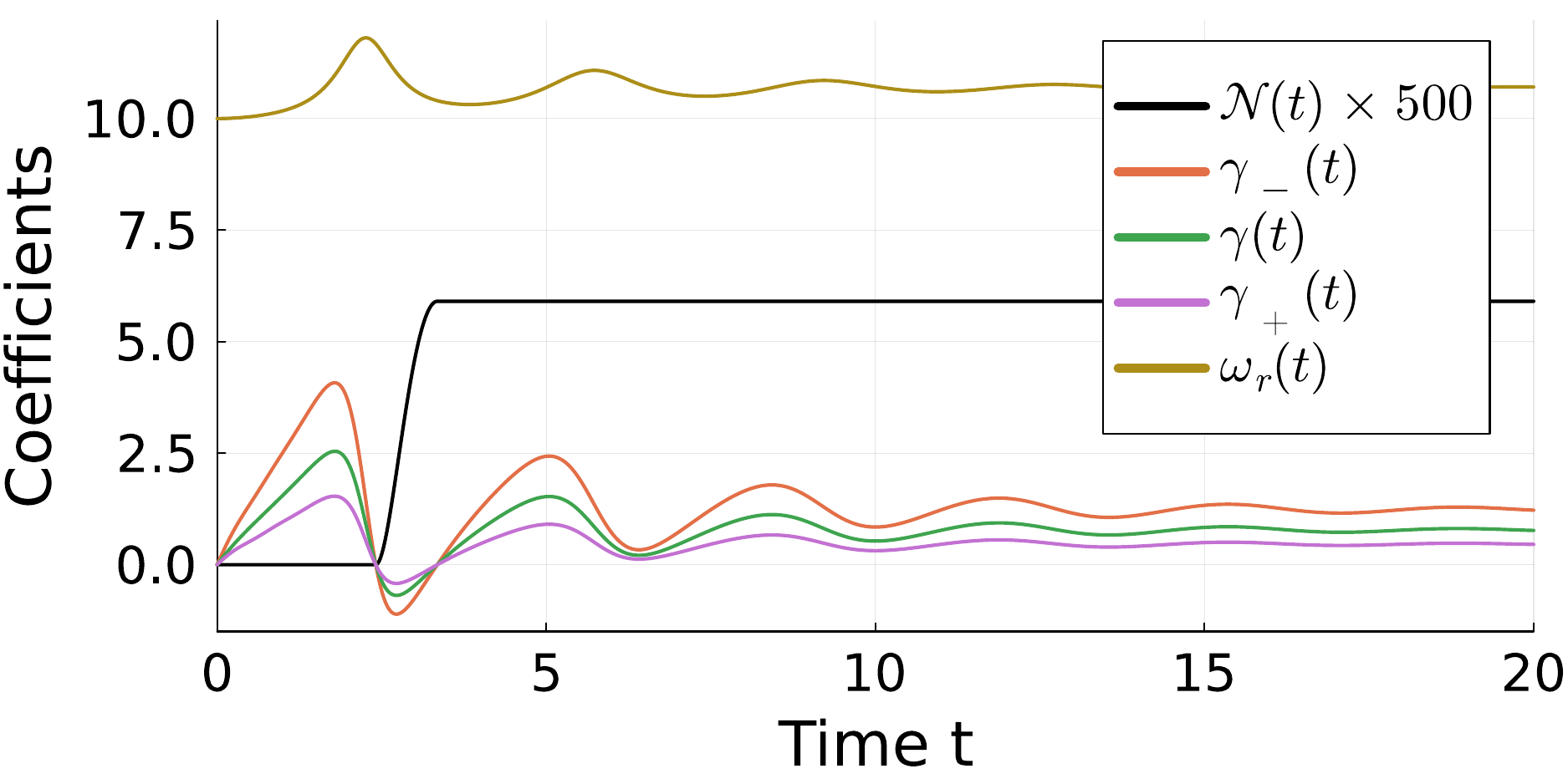}
	\caption{Exact coefficients and non-Markovianity for the state evolution with $\gamma_0 = 2.0$ and 
	$\Delta = 0.4$.
	The figure shows that information backflow is associated with at least one negative rate in the dissipator of the master 
	equation.}
	\label{fig:NormalizedParametersNM_strong}
\end{figure}

Comparing the exact rates $\gamma_{\pm}$ with the increase of non-Markovianity in 
\Cref{fig:NormalizedParametersNM_strong}, we observe that only when at least one of the rates
becomes negative the non-Markovianity increases. Thus, non-Markovianity is connected here to a violation of
CP divisibility of the dynamical map.\cite{Rivas2010a,Rivas2014a} 
We note that the coefficients do not change sign
at exactly the same time as can be shown using Eqs.~\eqref{eq:FAME}, \eqref{eq:CoeffExactGam} and
\eqref{eq:CoeffExactGamP}.

\begin{figure}[htb]
	\centering
	\includegraphics[width=0.9\columnwidth]{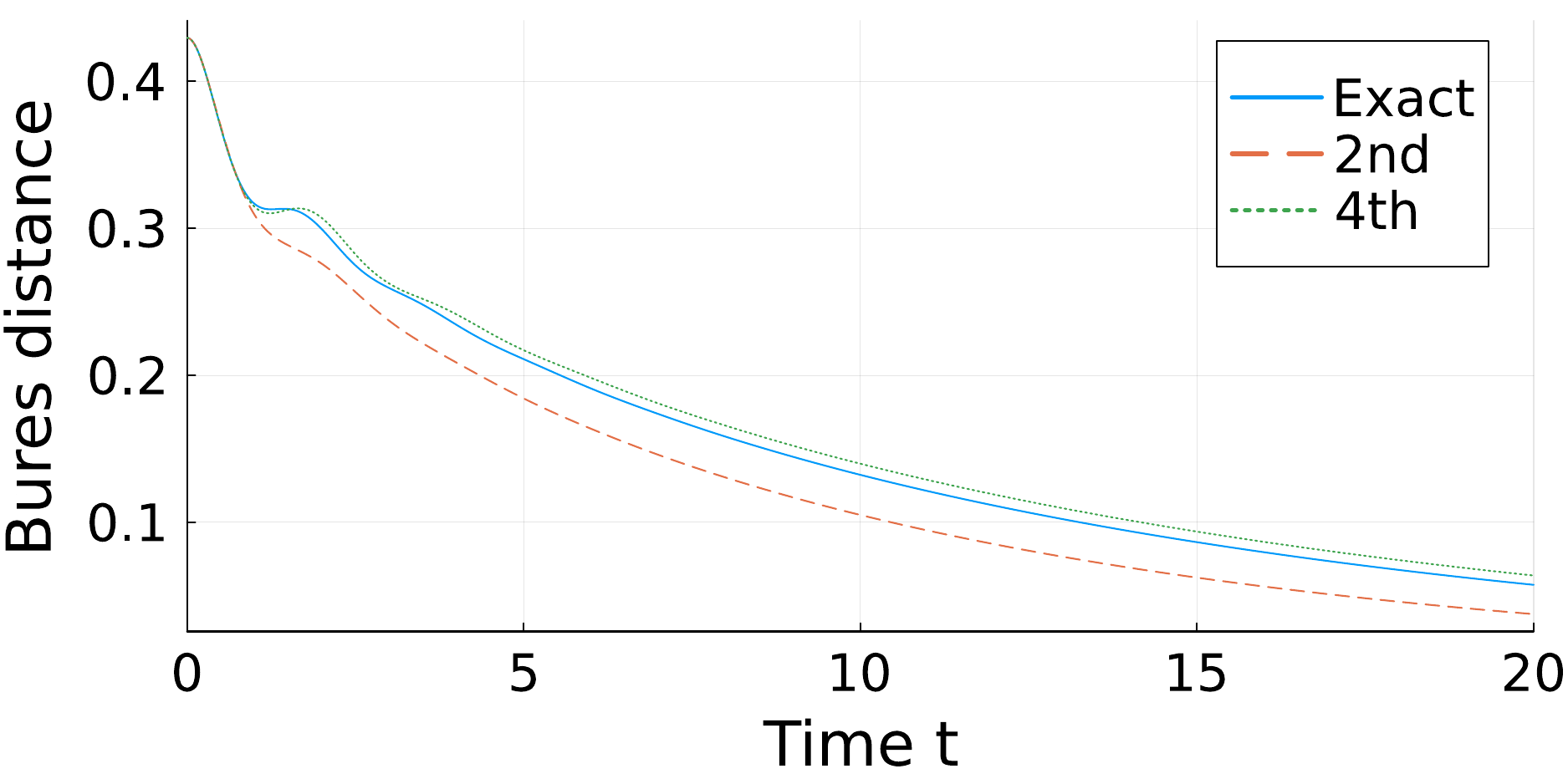}
	\caption{Two state Bures distance, computed exactly and perturbatively for $\gamma_0 = 2.0$ and detuning $\Delta = -3.0$. While the second-order approximation fails to reproduce the non-Markovian behavior, the fourth-order expansion captures the qualitative features of the non-Markovian dynamics.}
	\label{fig:Bures_strong_detuned}
\end{figure}

Increasing the detuning to $\Delta = -3.0$ enlarges the radius of convergence, and for the chosen coupling strength of $
\gamma_0 = 2.0$ the perturbative treatment becomes applicable again. For these values of the spectral density 
parameters, the resulting time evolution of the Bures distance is shown in \Cref{fig:Bures_strong_detuned}. The 
comparison demonstrates that even in the strong-coupling regime the perturbative expansion accurately reproduces the 
system dynamics as long as the fourth order coupling terms are included. This confirms that a sufficiently large detuning 
can restore the validity of the perturbative description and allow a reliable characterization of non-Markovian behavior.

\begin{figure}[htb]
	\centering
	\includegraphics[width=0.9\columnwidth]{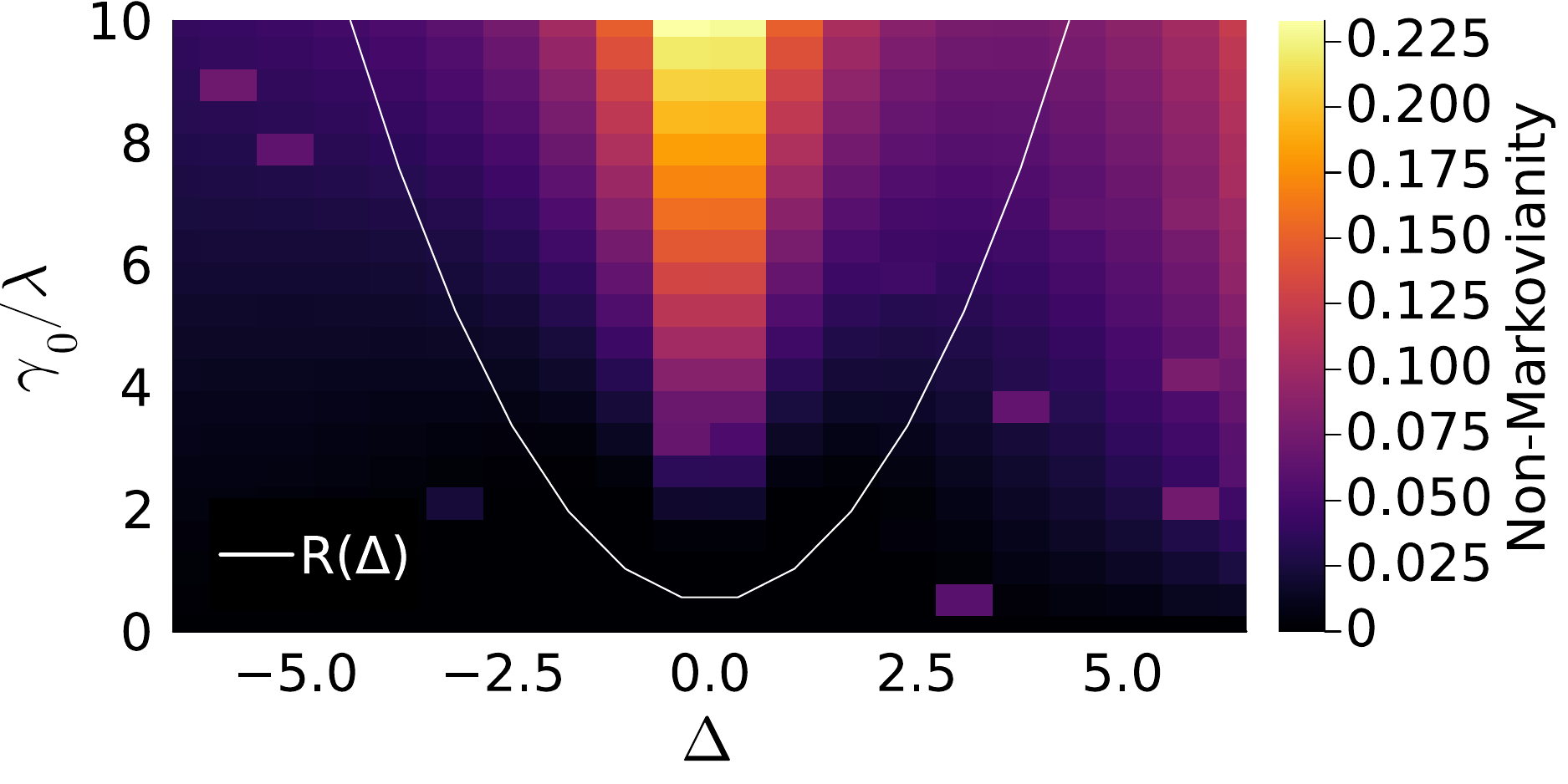}
	\caption{Non-Markovianity for various initial Lorentzian spectral densities, with the convergence radius shown in white, 
	determined from the exact dynamics.}
	\label{fig:Heatmap_analytical}
\end{figure}

\begin{figure}[htb]
	\centering
	\includegraphics[width=0.9\columnwidth]{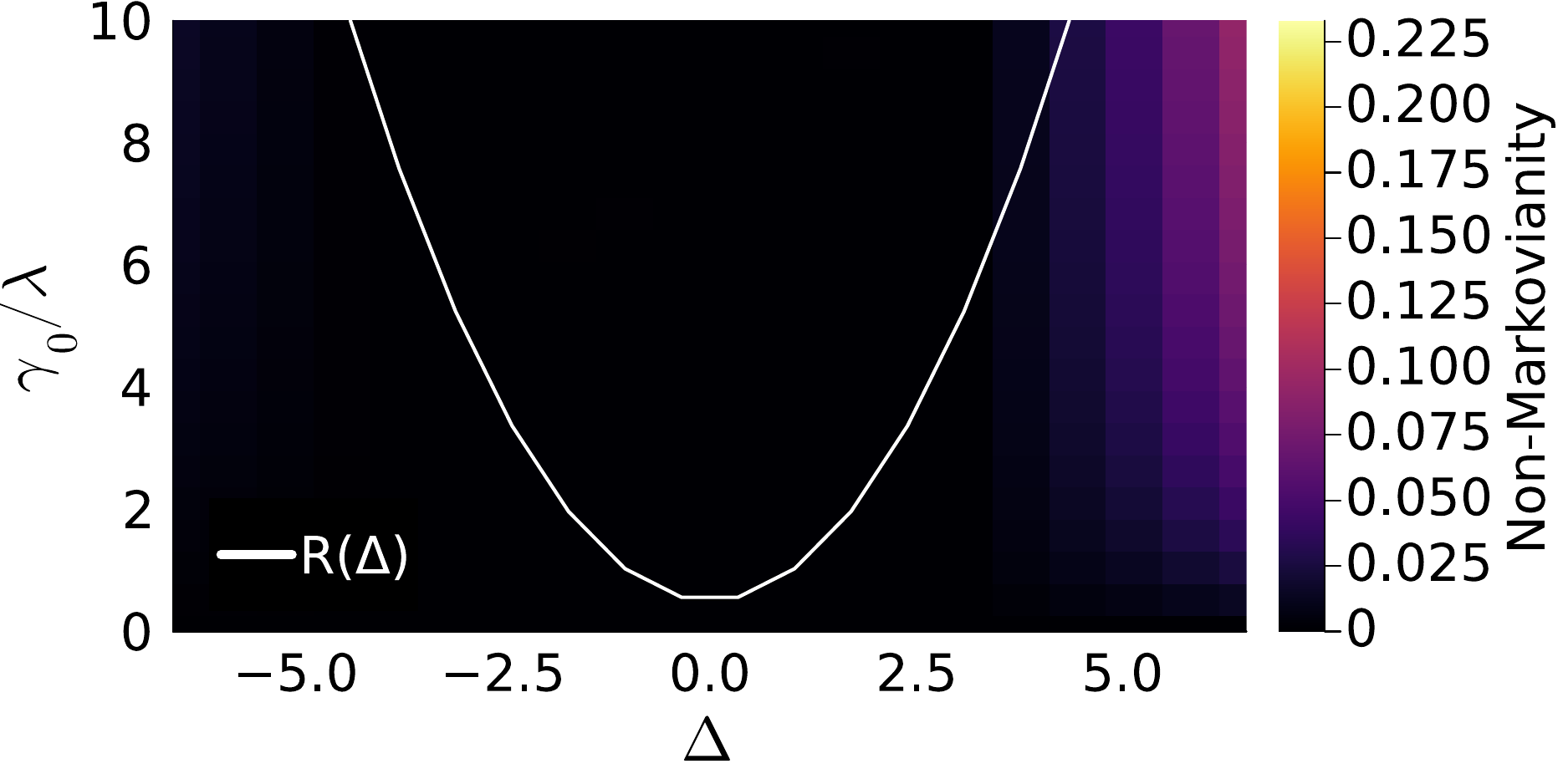}
	\caption{Non-Markovianity for various initial Lorentzian spectral densities, with the convergence radius shown in white, computed using the second-order approximation. The figure highlights the temperature-dependent asymmetry in detuning and the inability of the perturbative approach to capture the full parameter space.}
	\label{fig:Heatmap_2nd}
\end{figure}

\begin{figure}[htb]
	\centering
	\includegraphics[width=0.9\columnwidth]{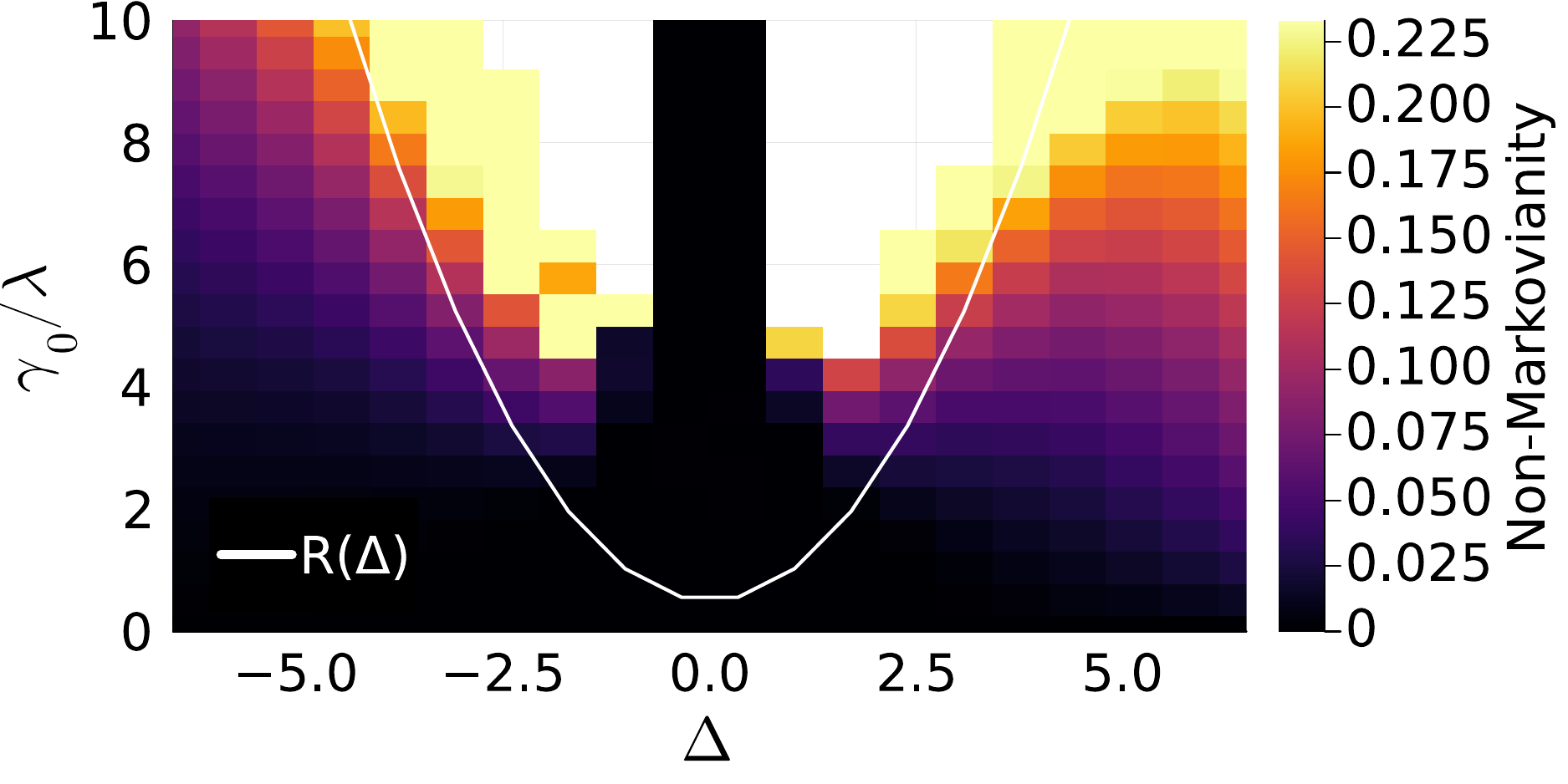}
	\caption{Fourth-order TCL approximation of the non-Markovianity for various initial Lorentzian spectral densities, with 
	the convergence radius in white, showing a more accurate representation of the full perturbative parameter space 
	(outer area).}
	\label{fig:Heatmap_4th}
\end{figure}

In \Cref{fig:Heatmap_analytical} we show the non-Markovianity measure \eqref{eq:NM} for the exact dynamics
as a function of the detuning  $\Delta$ and the coupling strength $\gamma_0/\lambda$, with a line indicating the radius of 
convergence $R(\Delta)$.
On the one hand, we can see that the non-Markovianity is not symmetric in detuning. This effect arises from the 
influence of temperature and the asymmetry of the terms $J(\omega)n(\omega)$; for zero temperature the 
non-Markovianity becomes symmetric with respect to detuning. 
Furthermore we can see that there are perturbative regimes where non-Markovianity is present 
and should be identifiable by the approximate approach.  
To examine this we calculate the same heat map and plot the non-Markovianity obtained by using the second and fourth 
order approximation in \Cref{fig:Heatmap_2nd,fig:Heatmap_4th}. 
We can see, that outside the perturbation limit also the representation of non-Markovianity fails (inner area). 
In particular, the near resonant cases with highest non-Markovianity measures are not correctly reproduced by the 
expansion, or cannot be calculated numerically (indicated by the white areas). When the detuning is sufficiently 
large to compensate for strong coupling, the perturbative treatment successfully captures non-Markovian effects 
where, however, the non-Markovianity measure is different especially for positive detuning.

We observe that the second-order approximation already reflects the asymmetry due to temperature in the highly 
detuned regions. In contrast, the fourth-order approximation provides a much more faithful representation of the overall 
structure of the non-Markovianity measure across the entire perturbative parameter space. Even close to the convergence 
boundary where it may slightly overestimate the magnitude it continues to offer valuable insight into the system dynamics. 


\section{Conclusion} \label{sec:Conclusion}

In this work we have investigated the performance of the expansion of the time-convolutionless (TCL) master equation 
for the Fano-Anderson model with a Lorentzian spectral density. To this end, we have studied the time-dependent 
decay rates $\gamma_{\pm}(t)$ and the renormalized frequency $\omega_r(t)$ in second and fourth order of 
the system-environment coupling and compared with these quantities with the corresponding coefficients of the exact 
TCL master equation. The dimensionless expansion parameter $\alpha^2$ is given by the ratio of the Markovian relaxation 
$\gamma_0$ to the width $\lambda$ of the spectral density, as one expects from the general 
theory.\cite{BreuerBook,Kampen} We have further identified the range of validity of this approach by determining the radius 
of convergence $R(\Delta/\lambda)$ of the Green function, which leads to a clear criterion for the applicability of the TCL 
expansion. 

To assess the physical implications of these results we have analyzed the impact on steady-state properties and on the 
characterization of non-Markovian behavior using the Bures distance. While the convergence radius imposes a 
fundamental bound on the perturbative treatment, our results show that  non-Markovian effects can still be reliably 
reflected within suitable parameter regimes. Even in the strong-coupling regime, the method offers a qualitatively accurate 
representation of the system dynamics and the associated measures of information backflow, determined by the Bures 
distance.

Moreover, our analysis highlights that detuning can effectively extend the perturbative regime by increasing the radius of 
convergence, thereby restoring the validity of the expansion even at coupling strengths where it would otherwise fail. Within 
this enlarged domain, the fourth-order TCL approximation reproduces both transient oscillatory features and long-time 
behavior with high accuracy, while the second-order expansion already captures key qualitative trends such as 
temperature-dependent asymmetries. These findings indicate that the perturbative TCL framework, when applied 
judiciously, provides a powerful and computationally efficient alternative to fully numerical approaches. The close 
agreement between our perturbative predictions and the exact solution in the regimes of convergence further underlines the 
broader applicability of higher-order TCL methods to the study of non-Markovian open quantum systems.



\section*{Acknowledgments}
We thank Irene Ada Picatoste and Alessandra Colla for many valuable insights, helpful conversations and 
comments on the manuscript. This work was supported by the German Research Foundation (DFG) 
through research unit \textit{Reducing Complexity of Nonequilibrium Systems} (FOR5099).

\section*{Data availability statement}
The data that supports the findings of this study are available within the article.

\bibliographystyle{apsrev4-2}
\bibliography{Lib.bib}
   	
\end{document}